\documentclass[%
 aip,
 amsmath,amssymb,
 reprint,%
]{revtex4-1}

\usepackage{graphicx}
\usepackage{dcolumn}
\usepackage{bm}

\usepackage[utf8]{inputenc}
\usepackage[T1]{fontenc}
\usepackage{mathptmx}
\usepackage{graphicx}
\usepackage{amsmath}
\usepackage{xcolor}
\usepackage{subcaption}

\font\grassettogreco=cmmib10
\font\scriptgrassettogreco=cmmib7
\font\scriptscriptgrassettogreco=cmmib10 at 5 truept
\font\sansserif=cmss10
\font\scriptsansserif=cmss10 at 7 truept
\font\scriptscriptsansserif=cmss10 at 5 truept
\def\bgr{\fam=13}
\def\ssm{\fam=14}
\input amssym.def
\input amssym.tex
\def \ebf{{\bf e}}

\def \vbf{{\bf v}}
\def \xbf{{\bf x}}

\def \wbf{{\bf w}}

\def \Xbf{{\bf X}}

\def \Dop{{\mathchardef\alpha="710B \ssm \char'104}}

\def \Iop{{\mathchardef\alpha="710B \ssm \char'111}}

\def \Lop{{\mathchardef\alpha="710B \ssm \char'114}}

\def \Rop{{\mathchardef\alpha="710B \ssm \char'122}}

\def \Top{{\mathchardef\alpha="710B \ssm \char'124}}

\def \Vop{{\mathchardef\alpha="710B \ssm \char'126}}
\def \Wop{{\mathchardef\alpha="710B \ssm \char'127}}

\def\Xibf{{\mathchardef\Xi="7104 \bgr \Xi}}

\def\nubf{{\mathchardef\nu="7117 \bgr \nu}}
\def\xibf{{\mathchardef\xi="7118 \bgr \xi}}

\def\taubf{{\mathchardef\tau="711C \bgr \tau}}

\def \Ecal{{\cal E}}

\def \eps{\epsilon}

  2
  3
 4
 5

\def \spa{\vskip    .3	  truecm \noindent }

\def\spi{\vskip .15	 truecm  \noindent}
\def \pan {\par \noindent}

   \newcount\Silicon \global\Silicon=1
   \newcount\PC \global\PC=2
   \newcount\parziale
   \def\Colori#1{\global\parziale=#1
                \ifnum\parziale=\Silicon
                    \input colors
                    \gdef\Color##1{\Black{##1}}                    
                \else\ifnum\parziale=\PC
                    \input colordvi
                    \gdef\textRGB##1{\textColor{##1 0.}}
                    \gdef\GrayA##1{\textGray{##1}}
                    \gdef\GrayB##1{\textGray{##1}}
                    \gdef\GrayC##1{\textGray{##1}}
                    \gdef\GrayD##1{\textGray{##1}}
                    \gdef\GrayE##1{\textGray{##1}}
                    \gdef\GrayF##1{\textGray{##1}}
                    \gdef\GrayG##1{\textGray{##1}}
                    \gdef\GrayH##1{\textGray{##1}}\fi                   
                \fi}        
\def\mean#1{\langle \,#1\,\rangle}

\def \parton#1{\left({{#1}}\right)}
\def \parqua#1{\left[{{#1}}\right]}

\def \derp#1#2{{\partial{#1} \over \partial{#2} }}

\font \rmsmm=cmr7
\def \hbix#1{\rmsmm \hbox{ {#1} } }

\def \underwrite#1{\mathop{\vtop{\ialign {##\crcr
$\hfil\displaystyle {#1}\hfil$\crcr\noalign{\kern3pt\nointerlineskip}
\crcr\noalign{\kern3pt}}}} \limits}

\def\Reali{\Bbb R}

\def\Toro{\Bbb T}

\def\sqr#1#2{{\vcenter{\hrule height.#2pt
     \hbox{\vrule width.#2pt height#1pt \hskip#1pt
       \vrule width.#2pt}
     \hrule height.#2pt}}}

\def\DAL{\hbox{\raise.250ex \hbox{$\sqr7{10}\,$}}} 

\def\bib#1.#2/#3/#4/#5/#6/#7.{\frenchspacing\item{[#1]}#2:\ {\it ``#3''}
~--~#4\ $\underline{\bf #5}$,\ #6 (#7)}
\def\diagramma#1#2#3#4#5#6#7#8{
 \vbox to 2.5cm{
       \hbox to 3cm{\hfil ${#1}$ \hfil}
       \hbox to 3cm{ ${#2}$\rightarrowfill ${#3}$ }
\hbox to 3cm{${#4} \Biggl\uparrow\hfill\Biggr\uparrow {#5}$}
       \hbox to 3cm{ ${#6}$\rightarrowfill ${#7}$ }
       \hbox to 3cm{\hfil ${#8}$ \hfil}   }    }
\def\sopra#1#2{{\raise 0.8 ex
\hbox{$
{{\scriptstyle \,{#2}}	\atop \displaystyle{#1}}
$}}
}

\def\figuraps#1#2#3#4#5{
\par
\midinsert
\centerline{\bf #4}
\vbox to #3 truecm{
\vskip #3 truecm
\ifnum #1 = 0	
\special {ps: plotfile #2}
\else		
\special {#2 0 0 moveto 16} \fi
}
\centerline{#5}
\endinsert}

\font\cofon = cmr6
\font\cobfon = cmbx6
\font\copi = cmr9
\def\codlib{{\copi\copyright}{\cofon 88-08- }{\cobfon 9820}}
\def\riga{\vskip .1  truecm   \hrule \vskip .2	    truecm \noindent }
\def\HeadLinea#1#2{	\headline={\vbox to 0pt{\vss\noindent
{\ifnum \pageno=1  \hfill {\bf \folio}		 
\else {\ifodd \pageno			   
{\noindent     \hfill  {\it     #2} \quad {\bf \folio}
 }\riga
\else				 
{\noindent {\bf\folio} \quad  {\it  #1} \hfill 	}
\riga
\fi } \fi }			}}
}
\def\testatina#1#2{	\headline={\vbox to 0pt{\vss\noindent
{\ifnum \pageno=1  \hfill {\bf \folio}		 
\else {\ifodd \pageno			   
{\noindent  \codlib   \hfill  {\it     #2} \quad {\bf \folio}
 }\riga
\else				 
{\noindent {\bf\folio} \quad  {\it  #1} \hfill \codlib	}
\riga
\fi } \fi }			}}
}
\def\testatinacap#1#2#3{	\headline={\vbox to 0pt{\vss\noindent
{\ifnum \pageno=#3  \hfill {\bf \folio}		 
\else {\ifodd \pageno			   
{\noindent  \codlib   \hfill  {\it     #2} \quad {\bf \folio}
 }\riga
\else				 
{\noindent {\bf\folio} \quad  {\it  #1} \hfill \codlib	}
\riga
\fi } \fi }			}}
}
\def\oggi{\number\day\space\ifcase\month
   \or gennaio\or febbraio\or marzo\or aprile\or maggio\or giugno\or
   luglio\or agosto\or settembre\or ottobre\or novembre\or dicembre
   \fi\space\number\year}
\def\today{\number\day\space\ifcase\month
   \or January\or February\or March\or April\or May\or June\or
   July\or August\or September\or October\or November \or December
   \fi\space\number\year}   
\def\frame#1{\ifmmode\dframe{#1}\else\leavevmode\lower 2.4 pt
    \hbox{\vrule\unskip\vbox{\hrule\kern 1.5 pt\hbox{\kern
    1.5 pt{#1}\kern 0.5 pt}\kern 2 pt\hrule}\unskip\vrule}\fi}

\def\dframe#1{\hbox{\vrule\unskip$\vcenter{\hrule\kern 3 pt\hbox
    {\kern 3 pt$\displaystyle{#1}$\kern3pt}\kern 3 pt\hrule}$\vrule}}

\def\eps{\epsilon}
\def\Tr{\,\hbox{Tr}\,}

\def\det{\,\hbox{det}}

\def\eps{\epsilon}
\def\taubf{\bf{\tau}}
\def\nubf{\bf{\nu}}
\def\xibf{\bf{\xi}}
\def\cot{\,\hbox{cot}}

\title{Propagation of rays in corrugated  waveguides: a stability analysis \\
with Lyapunov and Reversibility fast indicators.}
\begin{document}

\preprint{AIP/123-QED}

\title[]{Propagation of rays in 2D and 3D   waveguides: a stability analysis\\
with Lyapunov and Reversibility fast indicators}

\author{G. Gradoni}
 \email{gabriele.gradoni@nottingham.ac.uk.}
\affiliation{%
School of Mathematical Sciences and Department of Electrical and Electronic Engineering, University of Nottingham, United Kingdom
}%
\altaffiliation[Also at ]{Maxwell Centre, University of Cambridge, United Kingdom.}

\author{F. Panichi}%
 \email{federico.panichi@studio.unibo.it.}
\affiliation{ 
Department of Physics and Astronomy , University of Bologna, Italy
}%

\author{G. Turchetti}
 \email{giorgio.turchetti@unibo.it.}
\affiliation{%
Department of Physics and Astronomy, University of Bologna, Italy.   INDAM National Group of Mathematical Physics, Italy  
}%

\date{\today}

\begin{abstract}
Propagation of rays in 2D and 3D corrugated waveguides is performed in the general framework of stability indicators. The analysis of stability is based on the Lyapunov and Reversibility  error. It is found that the error growth follows a power law for regular orbits and  an exponential law for chaotic orbits. A relation with the Shannon channel capacity is devised and an approximate scaling law found for the capacity increase with the corrugation depth. 
\end{abstract}

\maketitle

\begin{quotation}
We investigate the propagation of a ray in a 2D wave guide  whose  boundaries are two parallel horizontal lines, with a periodic corrugation on  the upper line.
The reflection point abscissa  on the lower  line and the ray   horizontal velocity component after reflection are the phase space coordinates and the map connecting two consecutive reflections is symplectic.           
The dynamic behaviour is illustrated by the phase portraits which show that the regions of chaotic motion increase with the corrugation amplitude.
For a  3D wave-guide the 4D map connecting two consecutive reflections on the lower plane is symplectic, but its orbit cannot be examined by looking at the intersections with a 2D phase plane,  since a  continuous interpolation of the orbits is not  available.
In this case the  fast dynamic  indicators allow to perform a stability analysis.
For  each point of a grid in a 2D phase plane  one computes the orbit for a chosen number of iterations and the corresponding value of the fast indicator, which is conveniently visualized using a color plot.
After the fast Lyapunov indicator, many other  indicators have been introduced. Our analysis is  based on the
Lyapunov error (LE),  due to a small random initial displacement and the reversibility error occurring when  the orbit is reversed in presence of a small additive  noise   (RE) or round off (REM). 
For  integrable maps the growth of LE and RE follows a power law and for quasi integrable maps the same growth occurs  close to a stable fixed point.
More generally the error growth follows a power law for regular orbits and  an exponential law for chaotic orbits. There is numerical evidence that REM  grows as RE though with large fluctuations.  The  channel
capacity is related to LE  and the  dependence of its phase space average on the corrugation amplitude is considered. 
These indicators confirm their reliability for the stability analysis of the ray propagation in a 2D and 3D wave guide, providing a measure of the sensitivity of the orbits  to initial conditions, noise and round off.
\end{quotation}
%

%
%
%
\section{\label{sec:intro}Introduction}

\def\spi{\vskip 0.1 truecm \noindent}
The equivalence between geometrical optics and  mechanics was established in a
variational form by the principles of Fermat and Maupertuis.
If a ray propagates in  a uniform medium with a reflecting boundary,
then the trajectory is the same as a particle freely moving
and elastically colliding  with the same boundary.  As
consequence a wave-guide and a billiard 
are equivalent optical and mechanical systems \cite{ref0}. 
Since the velocity
of the particle does not change,  
we can assume it has unit modulus, just as the ray velocity normalized
to the speed of light.
The trajectory  is determined by the collision points and the velocity
direction  after the collision.
The billiards with polygonal or smooth convex closed  boundaries have
been intensively investigated and the mathematical literature is very rich,
see \cite{ref1,ref2,ref3,ref4}.
\spi
We first consider  the 2D wave-guide whose boundary is a  straight line  and a
corrugated parallel line.  The trajectory is a polygonal line specified by
the abscissa $x$ of the  reflection points on the straight line and the parallel component
$v_x$ of the velocity after reflection.
The Fermat principle establishes that, given two points  $x_1,x_2$ on the lower straight line,
the  ray,  colliding once with  the corrugated
line, follows the  path of minimal length and the collision is just a reflection.
In addition the ray  minimal path length $h(x_1,x_2)$  is the generating
function of the area preserving map $M$ connecting two subsequent reflections on the
straight line.
A similar procedure allows to obtain a symplectic map for the ray propagation
in a 3D waveguide made of an horizontal plane and an upper parallel
plane with a periodic corrugation.
If the medium between the waveguide boundaries if not uniform the   piecewise linear path   
joining two consecutive  reflection points on the lower line or plane is replaced by
the geodesic with respect to the metric $n\,ds$, where $n$ is the refraction index.
\spi
The  transition from ordered to chaotic motion
in  2D billiards and waveguides has been considered  \cite{ref5,ref6,ref7}; the transport and diffusion 
properties have been extensively analysed \cite{ref8,ref9,ref10,ref11,ref12,ref13,ref14}.
The stability properties of the map depend on the corrugation amplitude.
For the 2D wave guide the  phase portrait of the corresponding 2D map allows
to detect the regions of regular and chaotic motion. Finite time indicators
such as the Fast Lyapunov Indicator (FLI) \cite{Froeschle2000b,Froeschle2000a}
have been first proposed to analyze the orbital
stability. Other short term indicators of variational nature have been introduced:
the Alignment Indices  (SALI)
\cite{Skokos2001}, the Orthogonal Fast Lyapunov Indicator (OFLI)  \cite{Barrio2016},
the Mean Exponential Growth of Nearby Orbits (MEGNO) \cite{Cincotta2000,Cincotta2003},
which is an excellent  filter of  the oscillations of FLI, the relative Lyapunov indicator (RLI)
\cite{Sandor2004},  and the Generalized Lyapunov indicators (GALI), whose asymptotic behaviour is related  the  all the  Lyapunov exponents \cite{Bountis2008,Skokos20016}.  An indicator based on
the distribution of the stretching numbers (SSN) was also proposed \cite{Voglis1998,Voglis1999}.
An extensive numerical comparison of previous indicators for symplectic maps was 
presented \cite{Maffione2011,Maffione2012}.
Spectral indicators
based on the Fourier analysis have also been proposed and extensively used
\cite{Laskar1992}.  A geometric chaos indicator based on the Riemann curvature of the 
constant energy manifold \cite{Pettini1996,Pettini2008} and a 0-1 test of chaos 
\cite{Gottwald2005} have also been proposed. 
More recently the Lyapunov and reversibility errors have been
introduced \cite{Panichi2015, Panichi2017} to measure the sensitivity
of the orbits to a small random  initial  displacement  and  to  a small additive noise along
the orbit. The Lyapunov error (LE)
and the reversibility error
are simply related \cite{Turchetti-INTECH} so that the
computation of RE does not require expensive Monte-Carlo procedures.
The  relevant  difference of LE withe respect to previous variational indicators such as FLI,
is that LE  does not depend on the initial deviation  vectors.
A  full set of Lyapunov error indicators  (LEI), whose asyptotic behaviour is related to all the 
Lyapunov exponents, just as GALI,  and which  independent from the initial deviation  
vectors,    has been  introduced  jointly with  reversibility error  indicators (REI) \cite{Turchetti-PhysicaD}.   
The last  REI   corresponds to the  reversibility entropy,  and its 
asymptotic behaviour is governed by the sum of positive 
 Lyapunov exponents,  as the upper-bound to the  Kolmogorov-Sinali entropy \cite{KSentropy}.
 
The   LE and RE grow following a power law, for the regular orbits of quasi integrable  symplectic maps
as the previous variational indicators,  exponentially for chaotic orbits.
The reversibility error
due round off (REM) was first introduced in  \cite{Faranda2012} and its features
were examined in \cite{Zanlungo}. For previous works on the round
effect in the computation of orbits for Hamiltonian systems see \cite{Hairer2006}.
Previous numerical investigations of  Hamiltonian systems and symplectic maps  
confirm that LE   grows linearly  with oscillations,
(due to the  loss of rotational symmetry when the coordinates are not normal)
for regular orbits,  whereas the   growth of RE is almost oscillations free. 
Neglecting its large fluctuations  REM  is comparable with RE, 
even though no rigorous proof  is available. 
\spi
For the 3D waveguide no direct inspection of the orbits is possible, since
the section of the orbits of the symplectic 4D map with a 2D plane would 
require a continuous interpolation, 
available only when an  interpolating Hamiltonian is known.
Normal forms provide   the  interpolating Hamiltonian for
quasi integrable symplectic maps,
but  their recursive computation is possible just for polynomial maps,
and in addition the interpolation
is not exact due to the presence of a non integrable remainder.
The variational indicators,  computed
 for the orbits issued from the points of a regular grid in 
 a 2D  phase plane for  the 4D map, and visualized with a colour plot, allow to determine the 
stability properties just as  for
the 2D map.  We have analyzed only LE and RE for a limited number of iteration. Indeed 
as stated Froeschl\'e et al.  the variational indicators as FLI 
computed for short times exhibit  some dependency on the initial conditions 
of the deviation vectors.  Since  LE, RE and REM do not depend on the initial
deviation vectors  our choice is justified.  A careful investigation of the sticky chaos,
which   requires  longer orbits,
and   more extensive numerical exploration  of the reflection maps based 
on the full set  of invariant indicators  LEI and REI and  
a comparison with the  standard variational indicators,  will be the object of a future work.  
\\
Only  a few hundred iterations of the map  are required  to obtain a
reliable stability picture, unless one is interested in the details of  a small region.
To obtain the analytic form  of the tangent map is rather cumbersome for the 3D waveguide case.
The shadow orbit method provides a simple, though  less accurate,  alternative
which consists in   evaluating the orbits for initial
conditions with small displacements along an orthogonal basis, which amounts to replace the 
partial derivatives with finite differences. For any
initial condition, 4 additional evaluations of the orbit are required   (2 for the 2D map)
in order to obtain LE and RE, whereas    just 1 is required to compute REM.
\spi
We have analyzed a model of 2D waveguide for different values of
the corrugation amplitude,  showing that the LE, RE, REM provide comparable results, which
describe the orbits sensitivity to a small  initial  random displacement, to a noise
along the orbit and   to round off.
For  a model of  3D  waveguide  the same error plots, for initial conditions on  2D phase planes,
exhibit  a similar behaviour, though the structure is richer  with respect  to the 2D waveguide,
due to the presence of the Arnold web of resonances
The effectiveness of the proposed method, already experienced
in celestial mechanics \cite{Panichi2015,Panichi2017} and beam dynamics \cite{Arcidosso}
models,  is confirmed in these examples of 2D and 3D waveguides.

The paper is organised as follows. In section 1 we present the variational derivation of
the reflections map for a 2D waveguide. In section 3 the extension to the 3D waveguide is outlined. In section 4 
our dynamical indicators  are defined and their basic properties are  illustrated.
In section 5 and 6 the numerical results of  these  indicators  show how the 
instability regions grow when the corrugation amplitude increases. In section 7 the link
between our indicators and the channel capacity is established and it is shown how
its phase space average  increases with the corrugation amplitude. Conclusions and perspectives
are presented in section 8.
%
%
%
%
\section{The 2D waveguide}
Given two  parallel  reflecting lines $z=0$ and $z=1$ in the $(x,z)$ plane,
 the light  ray  direction  after each reflection on the lower line is  the same 
  being  specified by the unit vector $\vbf=(x_x,v_z=\sqrt{1-v_x^2})^T$, where
$T$ denotes the transpose of a matrix or a vector.  The time
$\tau$ between two reflection on the upper and lower line is given by $\tau v_z=1$.
We choose as phase space coordinates $(x,v_x)$ so that the sequence of reflections
$(x_n,v_{x\, n})$ on the lower line is given by
\begin{equation}
 \begin{split}
  v_{x\,n+1}&= v_{x\,n} \\
  x_{n+1} &= x_n+2\tau v_{x\,n} = x_n + \frac{2{v_{x\, n}}} {\sqrt{1-v^2_{x\,n}}}
 \end{split}
 \label{eq1}
\end{equation}
This map is integrable and area preserving. If we let $\vbf=(\cos \theta, \sin \theta)^T$ where
$0 \le \theta\le \pi$ the map becomes 
\begin{equation} \theta_{n+1}= \theta_{n}    \qquad \qquad
  x_{n+1} = x_n+\,2\cot \theta_n
 \label{eq2}
\end{equation}
and preserves the measure $d\mu=\sin \theta\, dx d\theta$. If our  phase space is 
a cylinder $\Toro([-\pi,\pi])]\times [-1,1]$  rather than the infinite strip $\Reali\times [-1,1]$,
then $x$ is an angle variable and $v_x$ an action variable.
\pan
The frequency  $\Omega= 2v_x(1-v_x^2)^{-1/2}$
diverges  for $v_x\to \pm 1$ and the interpolating Hamiltonian is $H= -2\sqrt{1-v_x^2}$.
Since also $\Omega'$ diverges when $v_x\to \pm 1$, any perturbation  renders the
map chaotic as for  an  integrable near the the separatrix where $\Omega$ vanishes but
its derivative diverges. 
The 2D waveguide is obtained by corrugating the upper line according to
\begin{equation}
 z=1+\eps f(x) 
\label{eq3}
\end{equation}
where $f(x)$ is a periodic function of period $2\pi$  such that $f(x)>-1$   and $ 0 \le \eps \le 1$.
The curvilinear abscissa  $s(x)$ of a point $Q$ on the corrugated line of coordinates $(x,z)$
where $z$ is given by (\ref{eq4})
\begin{equation}
 s(x)= \int _0^x\, \sqrt{1+\eps^2 {f'}^2(x')}\,dx'  
\label{eq4}
\end{equation}
Consider a ray which starts from  $P_0=(x_0,0)$  and reaches corrugated line at $Q=(x,z)$.
After reflection the ray reaches the $x$ axis at the point  $P_1=(x_1,0)$.
Keeping $P_0$ and $P_1$ fixed and letting $Q$ vary the path length $H$
of the segments $P_0Q$ and $QP_1$ is a function of $s(x)$
\begin{equation}
  \begin{split}
    H(s)&= h(x_0,s)+ h(x_1,s)  \\ 
    h(x_0,s)& = \sqrt{(x-x_0)^2 +(1+\eps f(x))^2} \qquad \\ 
    h(x_1,s) &= \sqrt{(x-x_1)^2 +(1+\eps f(x))^2} 
 \end{split}
  \label{eq5}
\end{equation}
where $x=x(s)$ is the inverse of the function $s=s(x)$ defined by (\ref{eq4}).
Referring to Fig. 1 we define $\psi$ and $\psi'$ the angles which  the velocities  $\vbf_0$
and $\vbf$ of the incoming and outgoing  ray at $Q$ forms with the tangent $\taubf$.
%
\begin{figure}[!t]
  \centering
  \includegraphics[width=0.5\textwidth]{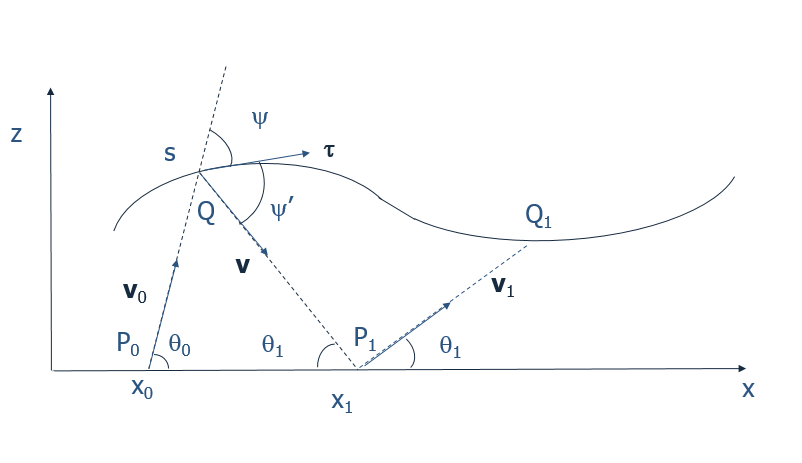}
\includegraphics[width=0.5\textwidth]{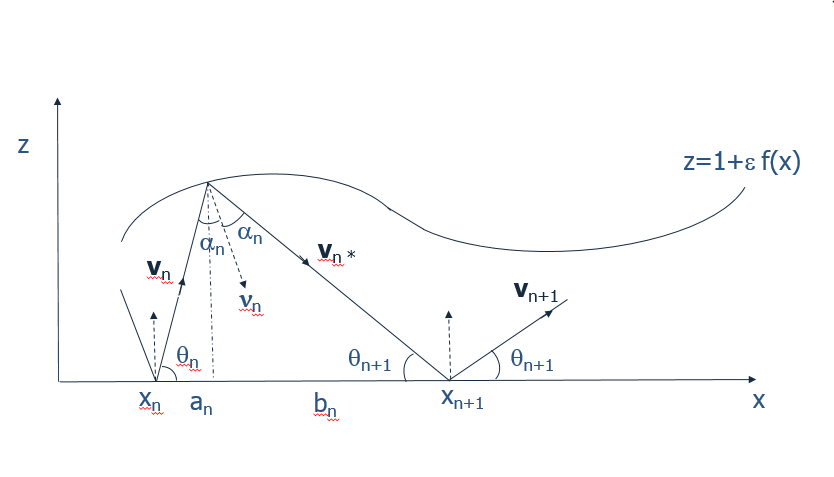}
\caption{Geometry of the reflections. On the right frame for $n=0$ we have $\alpha_0=\pi/2-\psi$, where $\psi$ is defined on the left frame.}
\label{fig:fig1}
\end{figure}
%
The angles between the vectors $-\vbf_0$ and $\vbf$ and  the normal $\nubf$ at $Q$  are $\pi/2-\psi$
and $\pi/2-\psi'$. As a consequence
\begin{equation}
  \begin{split}
    \vbf_0 & = {(x-x_0,\,\, 1+\eps f(x))^T\over h(x_0,s) }   \qquad
    \vbf = {(x_1-x,\,\, -1-\eps f(x))^T\over h(x_1,s) } 
    \\ \\
    \taubf& = {(1,\,\,\eps f'(x))^T\over \sqrt{1+\eps^2{f'}^2(x)} }  \qquad
    {\nubf} = { (\eps f'(x),\,\,-1)^T \over \sqrt{1+\eps^2{f'}^2(x)} } 
 \end{split}
\label{eq6}
\end{equation}
We denote with $\vbf_1$ the  velocity of the ray reflected at $P_1$ and with $\theta_0$ and $\theta_1$
the angles $\vbf_0$ and $\vbf_1$ form with the positive $x$ axis, see (\ref{fig:fig1}).
The derivatives of $h(x_1,s) $ and of $h(x_1,s) $ are given by
\begin{equation}
  \begin{split}
    \derp{}{s}\,h(x_0,s)&= {\,\,x-x_0+(1+\eps(f(s))\,\eps f'(x)\,\, \over h(x_0,s)\,\sqrt{1+\eps^2{f'}^2(x)} }=\\
    & \hskip 4.truecm =\vbf_0\cdot \taubf = \cos\psi   \\  \\  
     \derp{}{x_0}\,h(x_0,s)&= {x_0-x\over h(x_0,s)}=-v_{x\,0}=-\cos \theta_0
    \\ \\ \\
     \derp{}{s}\,h(x_1,s)&= {\,\, x-x_1+(1+\eps(f(s))\,\eps f'(x) \,\,\over h(x_1,s)\,\sqrt{1+\eps^2{f'}^2(x)} }= \\ 
     & \hskip 4 truecm =  \vbf\cdot \taubf = -\cos\psi'  \\ \\
     \derp{}{x_1}\,h(x_1,s)& = {x_1-x\over h(x_0,s)}=v_{x\,1}= \cos \theta_1
 \end{split}
\label{eq7}
\end{equation}
The stationary point  $H$ with respect to $s$, when $x_0$ and $x_1$ are kept fixed,
is met for $s=s_*(x_1,x_2)$ where $\psi=\psi'$,
which corresponds to the the reflection condition.
We introduce the function  $F(x_0,x_1)$ equal to $H$ at the stationary point
$s=s_*(x_0,x_1)$ and compute its differential 
\begin{equation}
  \begin{split}
     F(x_0,x_1) &= h(x_0,x_1,s_*) + h(x_0,x_1,s_*) \\ \\
     dF(x_0,x_1) &= -v_{x\,0}\,dx_0+ v_{x\,1}\,dx_1
  \end{split}
\label{eq8}
\end{equation}
Equation (\ref{eq8}) shows that $F(x_0,x_1)$ is  the generating function of the canonical
transformation $M$ from $(x_0,v_{x\,0})$ to $(x_1,\,v_{x\,1})$.
After $n$ iterations the phase space point $(x_n,v_{x\,n})$ is reached and is mapped into
$(x_{n+1}, v_{x\,n+1})$ by $M$. In the physical  space the ray issued from the point $P_n=(x_n,\,0)$
with horizontal velocity $v_{x\,n}$ is reflected at $Q_n=(x_n,1+\eps f(x_n))$ and reaches $P_{n+1}=(x_{n+1},\,0)$
where the horizontal velocity after reflection is $v_{x\,n+1}$ see figure \ref{fig:fig1}.

In order to obtain the map  we first  compute the time $\tau$ needed
for the transit between $P_n$ and  $Q_n$.
Then we determine the horizontal component of the velocity $v_{x\,n+1}$ of the ray
outgoing from $P_{n+1}$  which is equal to $v_{x\,*}$,  where the velocity $\vbf_*$
of the ray  from  $Q_n$ to  $P_{n+1}$ is determined by the reflection
condition $\vbf_*= \vbf_n-2\,\nubf\,(\nubf\cdot\vbf_n)$  and 
the normal of the guide at  $Q_n$ is given by
(\ref{eq6}) . Finally $x_{n+1}-x_n$ is given by $v_{x\,n}\tau+v_{x\,n+1}\tau'$
where the transit time from $\tau'$ from $Q_n$ to $P_{n+1}$ is given by
$v_{z\,n+1}\,\tau'=v_{z\, n}\tau$. 
\spa
\begin{equation}
  \begin{split}
    &  v_{z\,n } \tau = 1 +\eps f(x_n+ \tau v_{x\,n} ), \qquad \quad 
    v_{z\,n}  = \sqrt{1-v_{x\,n}^2}  \\ \\
    & v_{x\,n+1} = v_{x\,n} -2 \nu_x (\nubf\cdot\vbf_n)= \\ 
    & \hskip 3.6 truecm =v_{x\,n}-2{\eps^2\,{f'}^2\,v_{x\,n}
    - \eps\,f'\,v_{z\,n} \over 1+ \eps^2\,{f'}^2}   \\  
    & f'  = f'(x_n+\tau\,v_{x\,n}) \\  \\
   & x_{n+1}  = x_n+ \tau\,\parton{ v_{x\,n}\,+\,v_{x\,n+1} \,\,{v_{z\,n}\over v_{z\,n+1} } } \\ 
   & \hskip 4.5 truecm  v_{z\,n+1} = \sqrt{1-v_{x\,n+1}^2}
 \end{split}\
\label{eq9}
\end{equation}
The map $(x_{n+1},\,v_{x\,n+1})= M(x_n,\,v_{x\,n})$  is symplectic since it is implicitly
defined by the generating function $F(x_n,\,x_{n+1})$ according to
$v_{x\,n}=-\partial F/\partial x_n$ and $v_{x\,n+1}=\partial F/\partial x_{n+1}$.
The computation of the tangent map,  is given in the Appendix.
%
%
%
%
\section{The 3D wave-guide}
In this case we consider two planes $z=0$ and $z=1$ and the ray velocity
is the unit vector $\vbf$ which can be written in Cartesian coordinates according to
\begin{equation}
  \begin{split}
  \vbf&=(v_x,\,v_y,\,v_z) \qquad \qquad v_z= \sqrt{1-v_x^2+v_y^2}   
  \end{split}
\label{eq10}
\end{equation}
With  $\vbf$ we denote the velocity of the ray reflected by the lower  plane $z=0$ so that $v_z>0$.
The time $\tau$ between a reflection on the lower and upper plane is $\tau v_z=1$ so that,
choosing $(x,y,v_x,v_y)$ as phase space coordinates, the map between two consecutive
reflections reads 
\begin{equation}
   \begin{split}
     v_{x\,n+1} &= v_{x\, n} \\ \\ 
     v_{y\,n+1} &= v_{y\, n} \\ \\ 
     x_{n+1} &= x_n +2\tau v_{x,n} = x_n+2 {v_{x\,n} \over v_z\,n} \\ y_{n+1} &= y_n +2\tau v_{y,n} =
     x_n+2 {v_{y\,n} \over   v_z\,n}
  \end{split}
\label{eq11}
\end{equation}
If  the map is defined on $\Toro^2([-\pi,\pi])\times[-1,1]^2$,   then 
$v_v,\,v_y$ are the  action  variables and the interpolating Hamiltonian is
$H=-2 \sqrt{1-v_x^2-v_y^2}$.
In the 3D waveguide  the  plane $z=1$ is corrugated, namely it is replaced by the surface 
\begin{equation}
 z=z(x,y)= 1+\eps \,f(x,y) 
\label{eq12}
\end{equation}
where $f(x,y)$ is a periodic function in $x$ and $y$ with period $2\pi$ and $ f>-1$ with
$0<\eps< 1$. As a consequence the map describing the rays propagation can be  defined on
$\Toro^2([-\pi,\pi])\times[-1,1]^2$. As in the 2D case we consider the sequence of points
$P_0,P_1,\ldots,P_n, \ldots$  on the $z=0$ plane. We denote by $Q_n$ the point
on the upper corrugated plane  hit by the ray  issued from $P_n$,  which after reflection
reaches the $z=0$ plane at $P_{n+1}$.  The sequence points $P_0,P'_1,\ldots,P'_n, \ldots$
on the torus  $\Toro^2$ is simply obtained taking the modulus with respect to $2\pi$
so that $(x'_n,y'_n)\in [-\pi,\pi]$.  
Letting $\nubf(x,y)$ be the normal at the corrugated surface at the reflection
point  $Q=(x,y,z)=1+\eps\,f(x,y)$, which explicitly reads
\begin{equation}
 {\nubf}(x,y)=  { (\eps\,f_x,\,\eps\,f_y,\, -1)^T\over \sqrt{1+\eps^2(f^2_x+f^2_y) }}
  \label{eq13}
\end{equation}
The  map $(x_{n+1},\,y_{n+1}, v_{x\, n+1},v_{y\,n+1})=
M(x_n,\,y_n, v_{x\, n},v_{y\,n})$  specifying the ray trajectory is given  by 
\begin{equation}
  \begin{split}
    &\tau\,v_{z\,n} = 1+\eps f(x_n+\,\tau v_{x\,n},\,y_n+\tau\, v_{y\,n}) \\ 
    & \hskip 5 truecm  v_{z\,n}=  \sqrt{1-v_{x\,n}^2-v_{y\,n}^2} \\   \\ 
    & v_{x\,n+1} = v_{x\,n}   -2 \,\nu_x\,({\nubf}\cdot \vbf_n) \qquad
    {\nubf}={\nubf}(x_n+ \tau\,v_{x\,n},  y_n+ \tau\,v_{y\,n}) \\ \\ 
    &v_{y\,n+1}  = v_{y\,n}   -2 \,\nu_y\,(\nubf\cdot \vbf_n)  \\  \\
    & x_{n+1} = x_n+\tau \,\parton{ v_{x\,n} +v_{x\,n+1}\, { v_{z\,n}\over v_{z\,n+1}} }  \\ \\
    &y_{n+1} = y_n+\tau \,\parton{ v_{y\,n} +  v_{y\,n+1}\, {v_{z\,n}\over v_{z\,n+1}  }} \\ 
    &\hskip  4 truecm  v_{z\,n+1}= \sqrt{1-v_{x\,n+1}^2-v_{y\,n+1}^2}
\end{split}
\label{eq14}
\end{equation}
The first equation implicitly defines the propagation time $\tau$ from $P_n$ to $Q_n$,
the remaining  equations  define the map $M$ which is symplectic. Indeed  starting
from the Fermat variational principle it is shown in appendix B that this map
is obtained from a generating function $F(x_n,y_n,x_{n+1},y_{n+1})$. We do not quote in
this case the expression of the tangent map since it is rather involved.
%
%
%
%
\section{Dynamical indicators}
We will present a numerical analysis of the 2D and 4D
symplectic maps   whose orbits describe the propagation
of a ray in a 2D and 3D  periodic waveguide.
The stability analysis is intended to discriminate the regions
of regular and chaotic motions using the recently introduced
fast  stability indicators, denominated  Lyapunov and Reversibility error.
The first one is closely related to  fast Lyapunov indicator first proposed
to analyze the growth of a small initial displacement.
We shall not present a comparison  with other fast indicators,
because  it has been performed for other  models and since our indicators are 
independent from  the initial displacement.
\subsection{Lyapounov error}
For a symplectic map
in a phase space of dimension $2d$ map the Lyapunov error (LE)
describes the growth of an initial random displacement
$\eps\,\xibf $ where $\xibf \in \Reali^{2d}$ is a random vector
with zero  man  and unit variance
\begin{equation}
  \mean{{\xibf}}=0 \qquad  \qquad \mean{{\xibf}\,{\xibf}^T}= \Iop
  \label{eq15}
\end{equation}
in the zero amplitude limit $\eps\to 0$. Denoting with $\Xibf_n$
the random displacement after $n$ iteration and with $DM(x)$ the tangent map 
\begin{equation}
  \begin{split}
\Xi_n(\xbf)&= \lim_{\eps \to 0 } \, {M^n(\xbf+\eps{\xibf}) -M^n(\xbf) \over \eps }=
\Lop_n(\xbf) \,\xibf \\  \\
\Lop_n(\xbf)&=  DM^n(\xbf)=\Dop M(\xbf_{n-1})\, \Lop_{n-1}(\xbf)
\end{split}
\label{eq16}
\end{equation}
The square of  the Lyapunov error  $E_{L\,n}(\xbf)$ is defined as   as  the variance of
of the random vector $\Xi_n$ or  the trace  of  its    covariance matrix   
\begin{equation}
E^2_{L\,n}(\xbf) =  \mean {\Xi_n\cdot\Xi_n}= \Tr(\Sigma_n^2) \qquad \qquad
\Sigma_n^2= \mean {\Xi_n\,\Xi_n^T}= \Lop_n\,\Lop_n^T  
 \label{eq17}
\end{equation}  
We might define  the  error $E_{L\,n}(\xbf,\eta)$  for a given initial displacement
$\eps\eta$ with $\Vert \eta \Vert=1$,  when $\eps \to 0$. 
Its  logarithm  is just the Fast Lyapunov Indicator.
We observe that letting $\ebf_j$ be any orthonormal base 
the sum  of  $E^2_{L\,n}(\xbf,\ebf_j)$   extended to all the vectors of base is equal to
$E^2_{L\,n}(\xbf)$. Our definition is independent from the initial displacement.
An expression for  the Lyapunov error equivalent to (\ref{eq17}) is given by
\begin{equation}
 E_{L\,n}^2(\xbf)= \Tr(\Lop_n^T(\xbf)\,\Lop_n(\xbf))= \Tr\Bigl (\,(DM^n(\xbf))^T\,DM^n(\xbf)\,\Bigr)
  \label{eq18}
\end{equation}
 The stability analysis is performed  by fixing
$n$ and observing the change of  $E_n^L$ when $\xbf$ varies on a $2D$
phase plane (or a 2D manifold).
Oseledet theorem  \cite{Oseledet}  states that if $\xbf$ belongs to an ergodic component then
$(\Lop_n^T\,\Lop_n)^{1/2n}$ has a limit $\Wop\,e^\Lambda\,\Wop^T$  independent from $\xbf$,
where $\Wop$ is an orthogonal matrix,  $\Lambda$ is diagonal and its 
entries $\lambda_1\ge\lambda_2\ge\ldots \ge \lambda_{2d}$  are the Lyapunov exponents with 
$\lambda_{d+k}=\,-\,\lambda_{d-k+1}$ and $\lambda_k\ge 0$ for $1\le k\le d $.
As a consequence asymptotically $E^2_{L\, n} \,\simeq \,e^{2n\lambda_1 } \,+\, e^{2n\lambda_2 } \,+\,\ldots
+e^{2n\lambda_{2d}}\,\simeq \,e^{2n\lambda_1}$ and more rigorously $\lim_{n\to \infty}\,(E_{L\,n})^{1/n}= \lambda_1$.
However we  are not interested in the $n\to \infty$ limit but rather on
the dependence on $\xbf$ for a finite  (possibly large) value of $n$, when $\xbf$ 
varies in  phase space.
\subsection{Reversibility  error}
We consider now the iteration of the map  $n$ times followed  by the iterations of the inverse
map still $n$ times. Inserting  a small additive noise at each forward and backward
iteration,  the  final displacement with respect to the
initial condition is a stochastic process,  whose variance defines  the square on the reversibility
error (RE).  More precisely letting  $\xbf_{\eps, \,0}=\xbf$ be the initial conditions,
we consider the noisy
orbit 
\begin{equation}
  \xbf_{\eps, \,k}= M (\xbf_{\eps, \,k-1}) + \eps \xi_k  \qquad  \qquad
  \Xi_k= \lim_{\eps \to 0}\,{\xbf_{\eps, \,k}-\xbf_k \over \eps} 
\label{eq19}
\end{equation}
for $k=1,\ldots,n$.  Starting from $\xbf_{\eps,\, n}$ we consider the reversed noisy orbit
\begin{equation}
\begin{split}
  \xbf_{\eps,\, n,-k} &= M^{-1}(\xbf_{\eps, \,n-k+1})+ \eps \xibf_{-k}  \\ \\
  \Xi_{n,\, -k} &= \lim_{\eps \to 0}\,{\xbf_{\eps, \, n,-k}-\xbf_{n-k} \over \eps} 
\end{split}
\label{eq20}
\end{equation}
The random vector $\Xi_{k}$ satisfies a linear recurrence  with initial condition $\Xi_0={\bf 0}$
whereas $\Xi_{n,\,-k}$ satisfies another recurrence initialised by $\Xi_{n,\,0}=\Xi_{n} $,
see \cite{Panichi2017,Turchetti-INTECH} for an explicit expression.
We choose the random vectors $\xibf_{k}$  and 
$\xibf_{-k'}$  independent  for $k,k'>0$ 
\begin{equation}
\begin{split}
\mean { {\xibf}_k\,{\xibf}_{k'}^T} &= \Iop \,\delta_ {k,\,k'} \qquad 
\mean {{\xibf}_{-k}\,{\xibf}_{-k'}^T} = \Iop \,\delta_{k,\,k'} \\  \\ 
\mean {{\xibf}_{-k}\,{\xibf}_{k'}^T} &=
\mean {{\xibf}_{k}\,{\xibf}_{-k'}^T} = 0
\end{split}
\label{eq21}
\end{equation}
Letting $\Xi_{R\,n}= \Xi_{n,\,-n}$ be the stochastic displacement  with respect to the
initial condition $\xbf$ after reversing the orbit and $\Sigma^2_{R\,n}(\xbf) $ the corresponding covariance matrix
we define the reversibility error $E_{R\,}(\xbf)$ according to
\begin{equation}
\begin{split}
 E^2_{R\,n}(\xbf) &=  \mean {\Xi_{R\,n}\cdot\Xi_{R\,n}}= \Tr\,\bigl(\Sigma^2_{R\,n}(\xbf)\,\bigr) \\  \\
 \Sigma^2_{R\,n}(\xbf)  &= \mean {\Xi_{R\,n}\,(\Xi_{R\,n})^T} 
\end{split}
 \label{eq22}
\end{equation}  
The covariance matrix that $\Sigma^2_{R\,n}(\xbf)$ can be expressed in terms of the tangent
maps $DM^k(\xbf)$ and their
inverses $DM^{-k}(\xbf)$.  We do not quote their expression, which can be found in \cite{Panichi2015,Panichi2017}
since it can be  proved  \cite{Turchetti-PhysicaD}  that for a symplectic map $M$ in $\Reali^{2d}$ the error RE is related to LE by the following expression
\begin{equation}
 \begin{split}
  E_{R\,n}^2(\xbf) &=  E_{L\,0}^2(\xbf) + E_{L\,n}^2(\xbf)\,+\,  2\,\sum_{k=1}^{n-1}\, E_{L\,k}^2(\xbf) \\
  E_{L\,0}^2 &= \Tr(\Iop)=2d
  \end{split}
 \label{eq23}
\end{equation}  
or by the recurrence $ E^2_{R\,n}=E^2_{R\,n-1}+ E^2_{L\,n}+E^2_{L\,n-1}$ initialized by 
$E^2_{R\,0}=0$ and $E_{L\,0}^2=2d$. 
\\  \\
We finally define the reversibility error due to to round off (REM).
Letting $M_\eps(\xbf)$ the map computed
with a finite precision $\eps$ (typically in the 8 bytes representation of reals $\eps \simeq 10^{-17})$
and with $M^{-1}_\eps$ the inverse of $M$ computed with finite accuracy, so that 
$M^{-1}_\eps\bigl( M_\eps(\xbf)\bigr)\not = \xbf$,  the modified reversibility error (REM) is defined
according to 
\def\REM {{\hbix{REM}}}
\begin{equation}
E_{\REM\,n}(\xbf) = { \left \Vert M_\eps^{-n}\bigl(M_\eps^n(\xbf)\bigr) - \xbf\right \Vert  \over \eps}
 \label{eq24}
\end{equation}  
Letting $\xbf_{\eps,\,n}=M_{\eps}(\xbf_{\eps,\,n-1})$ be the orbit computed  with round off,
the local error $\bigl(\,M_{\eps}(\xbf_{\eps,\,n-1})- M(\xbf_{\eps,\,n-1})\,\bigr)/\eps$
is similar to a random vector, possibly correlated, if the map has a sufficiently
high computational complexity, see \cite{Hairer2006,Zanlungo}.
The difference with respect to the additive 
noise, we have considered to define RE,   is that  we have just a single realization.  As a consequence  even though
the behaviour of RE and REM is similar,  the last one exhibits large fluctuations, when $n$ varies.
On the other side the computation of  REM is really trivial since it  requires a  few lines of
code, if the inverse map is available. This is the case of the reflection map for a waveguide. Indeed  
to reverse the evolution after $n$ iterations  we must simply change the sign of 
the horizontal velocity components(s) and iterate again the same map $n$ times.
\subsection{Integrable and quasi-integrable maps}
The computation of LE and RE  can be analytically performed  for an integrable map.
If  the 2D map is a rotation $M(\xbf)= \Rop(\omega) \xbf$  then $E_{L\,n}^2=2$
and $E_{R\,n}^2=4n$. If the map  $M$  is linearly conjugated to a rotation by $\xbf=\Top \Xbf$
then the map in normal coordinates is  $\Rop(\omega) \Xbf$
and  $M(\xbf)= \Top\Rop(\omega)\Top^{-1} \xbf$ in the given coordinates. The loss of symmetry
in the coordinates $\xbf$ induces periodic oscillations $(E^2_{L\,n}=A+(2-A) \cos(\omega n)$ with
$A\ge 2$ and $E_{R\,n}$  has a linear growth with oscillations.
A map  $M(\xbf)$ is integrable if there is a coordinate  transformation $\xbf=T(\Xbf)$
such that the map becomes  $ N(\Xbf)= R(\Omega(J))\,\Xbf$  where the frequency depends on the action
$J={1\over 2}\Vert\Xbf\Vert^2$. In this case $M(\xbf)= T \circ N(\Omega(J))\circ \,T^{-1}(\xbf)$.
The Lyapunov and reversibility errors   LE and LE, depend on the initial condition and  in the normal
coordinates $\Xbf$ they grow according to
\begin{equation}
\begin{split}
  E_{L\, n}^2(\Xbf) &= 2+(2J\,\Omega'(J)\,n)^2 \\
  E_{R\,n}^2(\Xbf) & 4n +2\,\bigl(2J\Omega'(J)\bigr)^2\parton{ {n^3\over 3}+{n\over 6}} 
\end{split}
 \label{eq25}
\end{equation}  
The  Lyapunov error  $LE$  in the coordinates $\xbf$  exhibits a linear growth with
oscillations, due to the loss of rotational symmetry as in the linear case and we have
\def\ch{\hbox{ch}}
\def\sh{\hbox{sh}}
\begin{equation}
\begin{split}
 E_{L\,n}^2(\xbf)= \alpha_0(n)+ n\,J\,\Omega'(J) \, \alpha_1(n) \\ +\bigl(J\,\Omega'(J)\,n \bigr)^2\,
  \alpha_2(n) 
\end{split}
 \label{eq26}
\end{equation}   
where  $J={1\over 2}\Vert T^{-1}(\xbf)\Vert^2$ and $\alpha_k$ are periodic functions of $n$ with
frequency $\Omega(J)$. 
According to equation (\ref{eq23}) the  square of the reversibility error RE grows as $n^3$  with oscillations
whose amplitude  grows as $n^2$. 
If the map is quasi integrable as in the neighborhood of an elliptic fixed point at $\xbf_*$  with non-resonant
frequency,  then it  is conjugated with its normal   form  up to a remainder  which can be made exponentially small
with  $\sim e^{-a/r^\alpha}$ where  $r=\Vert \xbf-\xbf_*\Vert$.
Exponentially small  estimates  for the remainder  can  also  be obtained  for the error on the tangent map.  
\\
If the system has an hyperbolic point at $\xbf=(x,p)^T={\bf 0}$  and it is linear then
$M(\xbf)=(e^{-\lambda} x,\,e^\lambda p)^T$ and $E_{L\, n}^2=2\,\ch(2n\lambda)\sim e^{2n\lambda}$.
If the system is non linear,
in normal coordinates the map is $X_{n+1}= e^{-\lambda(XP)}X_n$ and $P_{n+1}= e^{\lambda(XY)}P_n$
with initial condition $\Xbf_0=(X,P)^T$. The Lyapunov error is
\begin{equation}
\begin{split}
  E_{L\,n}^2(\Xbf) = e^{2n\,\lambda}\Bigl( 1 + nXP\,\lambda'(XP)\Bigr)^2 
  \,\, \Bigl( 1+O\bigl(e^{-2\lambda\,n}\bigr) \,\Bigr) 
  \end{split}
 \label{eq27}
\end{equation}   
The square of LE   near  an  elliptic  equilibrium point for a generic symplectic map
grows linearly with oscillations,  the square of RE   grows as $n^3$ with
oscillations whose amplitude grows as $n^2$ so that asymptotically they become negligible.
Near a hyperbolic point  LE and RE  have  the same exponential growth.
 Finally REM  behaves just as  RE
but exhibits large fluctuations.  Right after the break up of
the invariant curves surrounding an elliptic point sticky orbits are found and  the indicators
have initially a power law growth followed by an exponential growth. We conclude  that
RE is the smoothest indicator,  since LE can exhibit oscillations  and REM is affected
by large fluctuations.  However  if we choose a fixed value of $n$ large enough, the
portraits all these indicators in a chosen plane of phase space are very similar.
%
%
%
\section{The  model of a 2D waveguide}
We present first a numerical analysis of the orbits for the map of the 2D
wave\-guide where the corrugation is given by 
\begin{equation}
   f(x)= \cos (x)
   \label{eq28}
\end{equation}
The  phase space coordinates are $(x,\,v_x)$  and the map
is defined on the cylinder $\Toro([-\pi,\pi])\times[-1,1]$  where $\Toro(([-\pi,\pi])$ is the interval
$[-\pi,\,\pi]$ with identified ends.
The map has an elliptic  fixed point at $x=0,\, v_x=0$  where $f(x)$ has a maximum
and  is approximated by $f(x) = 1-{1\over 2}x^2 +O(x^4)$. 
We expand the map retaining only the linear terms in $x,\,v_x$.   The first
equation in (\ref{eq9}) gives $\tau=1+\eps$   neglecting second order terms in $x$ and $v_x$.
Since  $f'(x)= - x + O(x^3)$
in the second equation the square of  $f'$ are neglected and $v_{z\,n}$ is replaced with 1.
Accordingly the second and third equations become
\begin{equation}
\begin{split}
  v_{x\,n+1} &=  v_{x\,n}  -2\eps\,(x_n+\tau v_{x\,n} )   \\ \quad x_{n+1} &= x_n + \tau(v_{x\,n}+v_{x\,n+1} )
\end{split}
   \label{eq29}
\end{equation}
Letting  $M(\xbf)=\Lop\xbf$  be the linear map we have
\begin{equation}
  \Lop= \begin{pmatrix} 1-2\eps \,\tau  & 2\tau -2\eps\,\tau^2 \\ \\
                        -2\,\eps   & 1-2\eps \,\tau    \end{pmatrix}
   \label{eq30}
\end{equation}
so that $\det \,\Lop=1$  and ${1\over 2}\,\Tr \,\Lop= 1-2\eps\,\tau$. The map is conjugated to a rotation
$\Lop=\Vop\Rop(\omega)\,\Vop^{-1}$  where $\sin^2(\omega/2)= \tau\eps=\eps(1+\eps)$.
In this case we have  $\Lop_n=\Lop^n$.
Close to $x=0$ the profile $z=1+\eps- \eps x^2/2$  is a  concave  mirror and the
elliptical trajectories in phase space $(x,v_x)$
correspond to caustics in  configuration space, namely the plane  $(x,z)$ where
the rays propagate.
Close to  $x=\pi$  the profile  is approximated by $z=1-\eps+{1\over 2}\, \eps\, (x-\pi)^2
+O(x-\pi)^4 $ so that $\tau=1-\eps$ and   the  linearized map  is 
$(x_{n+1}, v_{x\, n+1})^T=\Lop\,(x_n,\,v_{x\,n})^T$, where $\Lop$ is given by (\ref{eq29}) with
$\epsilon \to -\epsilon$. The map
$\Lop$ is   conjugated to a hyperbolic rotation $\Rop_H(\alpha)$ where
$\sh^2(\alpha/2)=\eps\tau=\eps(1-\eps)$.
\\ Near $x=\pi$ the wave guide corresponds to an    optical system  given by
a plane and  convex  mirror \cite{ref0}.
\\
The map  defined by (\ref{eq9}) was computed by solving the  equation for $\tau$
with  the bisection method initialized  by
$\tau_1= 0.5/\sqrt{1-v_{x\,n}^2}$ and $\tau_2= 1.5/\sqrt{1-v_{x\,n}^2}$.
The convergence is achieved  also when $v_{x\,n}$ approaches $1$ or $-1$.
The number of iterations
to reach machine accuracy varies  between 40 and 60. Newton's method is faster
and machine accuracy is
reached in less 10 iterations,  but convergence problems are
met when $v_{x\,n}$ approaches $1$ or $-1$.
%
\begin{figure}[!ht]
\centering
\begin{subfigure}{\columnwidth}
    \includegraphics[width=0.8\textwidth]{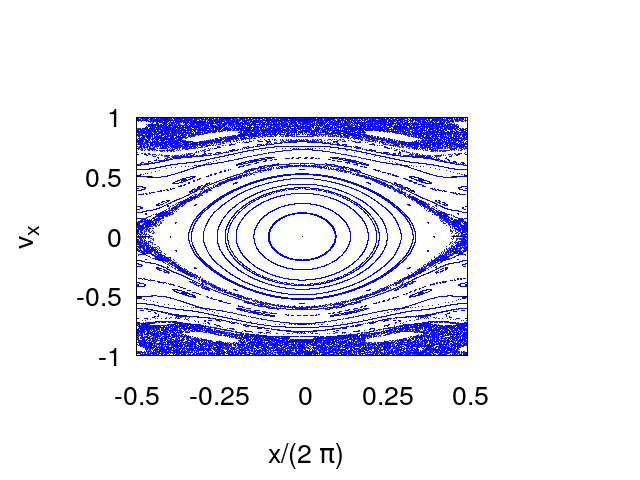}
    \caption{}
\end{subfigure}
\begin{subfigure}{\columnwidth}
    \includegraphics[width=0.8\textwidth]{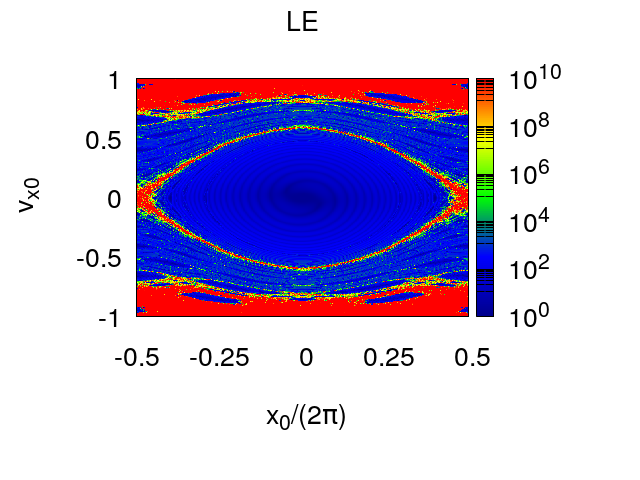}
    \caption{}
\end{subfigure}
\begin{subfigure}{\columnwidth}
    \includegraphics[width=0.8\textwidth]{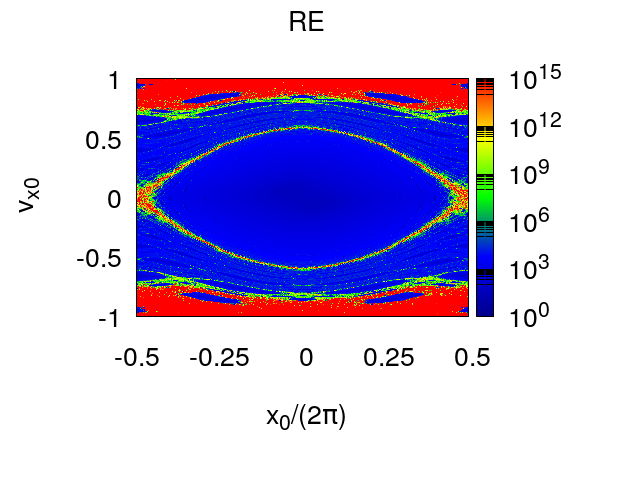}
    \caption{}
\end{subfigure}
\begin{subfigure}{\columnwidth}
    \includegraphics[width=0.8\textwidth]{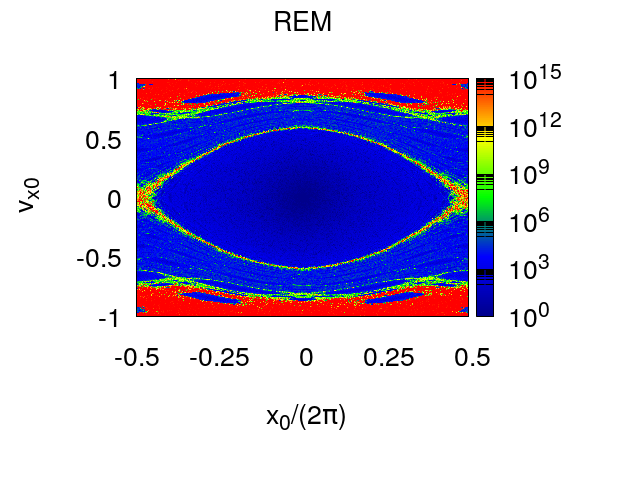}
    \caption{}
\end{subfigure}
\caption{From the top: (a) Phase portrait of the map for  the 2D waveguide with  corrugation $z=1+\eps\cos(x)$   and $\eps=0.1$.
(b) Colour plot of LE for $N=200$ iterations of the map.
(c) Plot of RE for  $N=200$ iterations of the map.
(d) Plot of REM for  $N=200$ iterations of the map.}
\label{fig:fig3} 
\end{figure}
We have also computed the tangent map and the explicit
expression is  written in  Appendix A. Writing the implicit equation $G(x_n,v_{x\,n},\tau)=0$,
the solution is not defined if $\partial G/\partial \tau=0$ a condition numerically never met.
In figures 2, 3 we compare the phase portraits with the
colour plots  of LE, RE and REM
for two different values of the corrugation amplitude $\eps=0.1,\,0.2$
respectively and orbits length $N=200$.  The correspondence is quite good and  RE appears to be
the smoothest indicator since it is free from oscillations and fluctuations when $n$ varies. 
Increasing $N$ does not change the plots significantly. Higher values of $N$ are needed if
one wishes to observe details in small regions where the transition from ordered to chaotic
orbits occurs.
\spa
\begin{figure}[!ht]
\centering
\begin{subfigure}{\columnwidth}
    \includegraphics[width=0.8\textwidth]{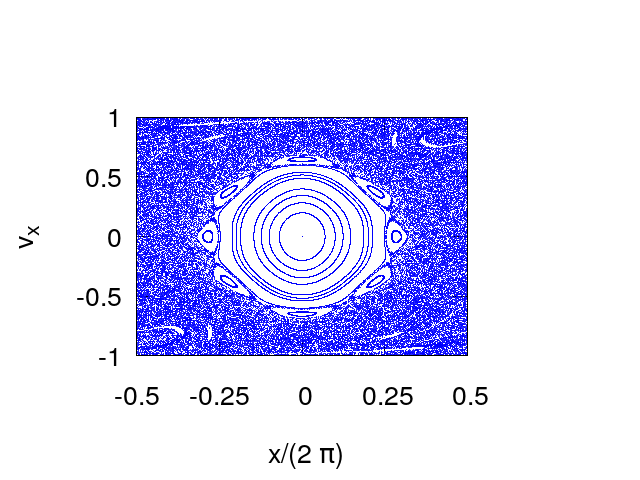}
    \caption{}
\end{subfigure}
\begin{subfigure}{\columnwidth}
    \includegraphics[width=0.8\textwidth]{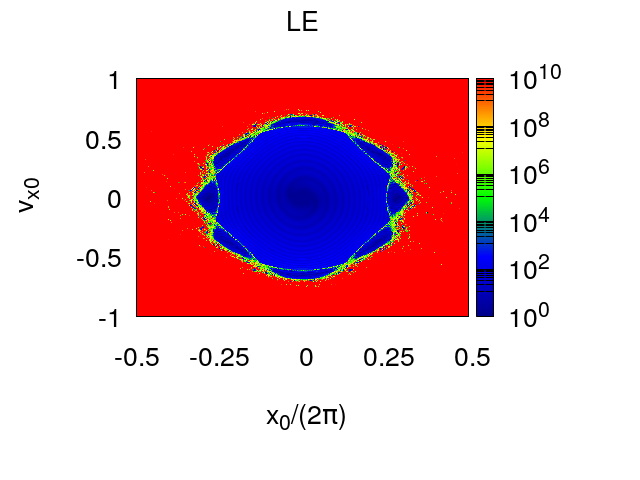}
    \caption{}
\end{subfigure}
\begin{subfigure}{\columnwidth}
    \includegraphics[width=0.8\textwidth]{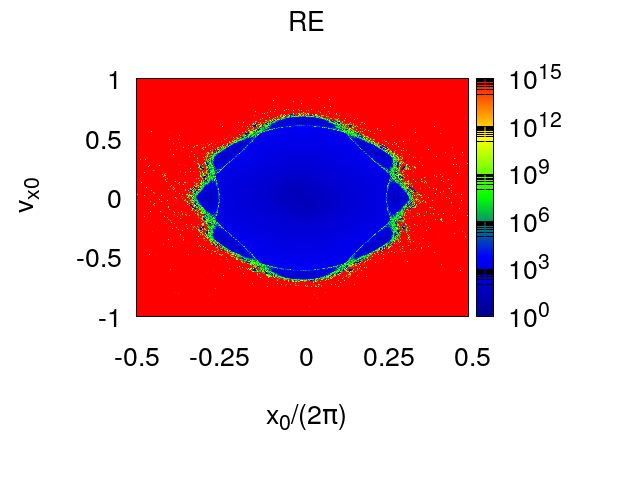}
    \caption{}
\end{subfigure}
\begin{subfigure}{\columnwidth}
    \includegraphics[width=0.8\textwidth]{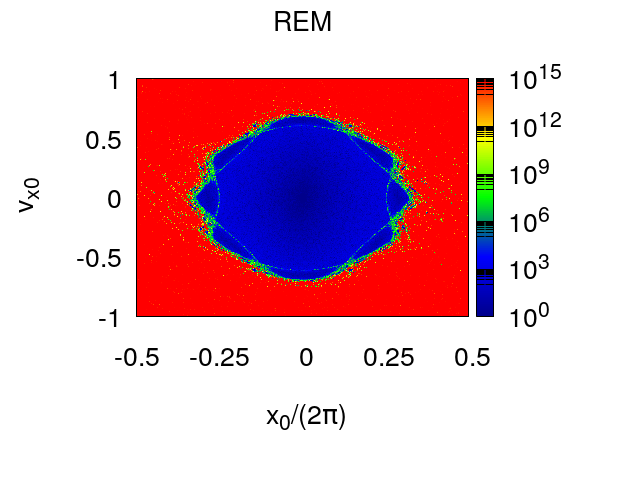}
    \caption{}
\end{subfigure}
\caption{From the top: (a) Phase portrait of the map for  the 2D waveguide with  corrugation $z=1+\eps\cos(x)$   and $\eps=0.2$.
(b) Colour plot of LE for $N=200$ iterations of the map.
(c) Plot of RE for  $N=200$ iterations of the map.
(d) Plot of REM for  $N=200$ iterations of the map.}
\label{fig:fig4} 
\end{figure}
%
%
%
%
%
%
%
%
%
%
%
%
%
\section{The 3D model}
\begin{figure}[!ht]
\centering
\begin{subfigure}{\columnwidth}
    \includegraphics[width=\textwidth]{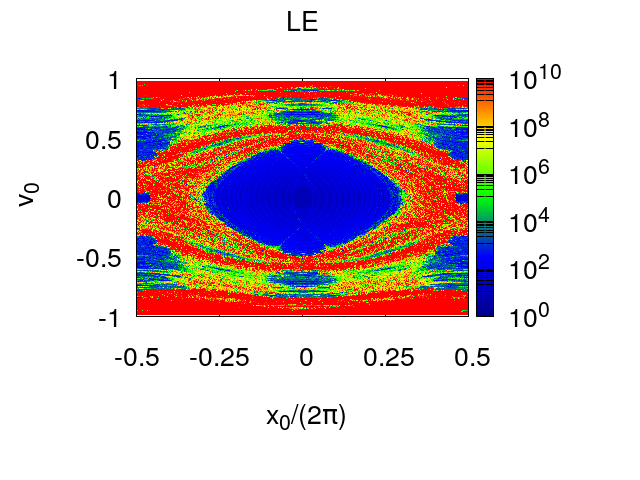}
    \caption{}
\end{subfigure}
\begin{subfigure}{\columnwidth}
    \includegraphics[width=\textwidth]{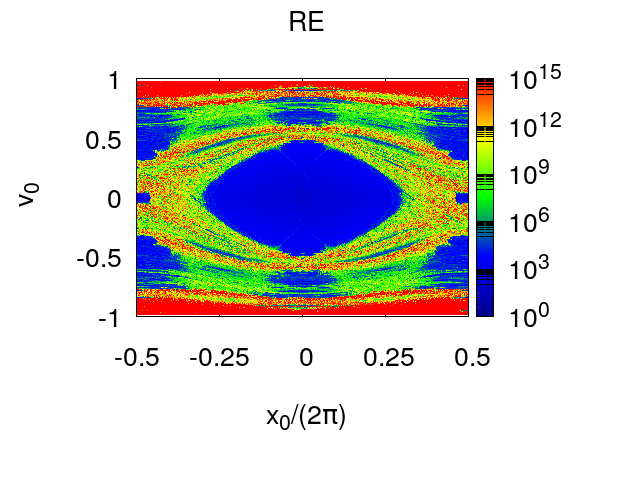}
    \caption{}
\end{subfigure}
\begin{subfigure}{\columnwidth}
    \includegraphics[width=\textwidth]{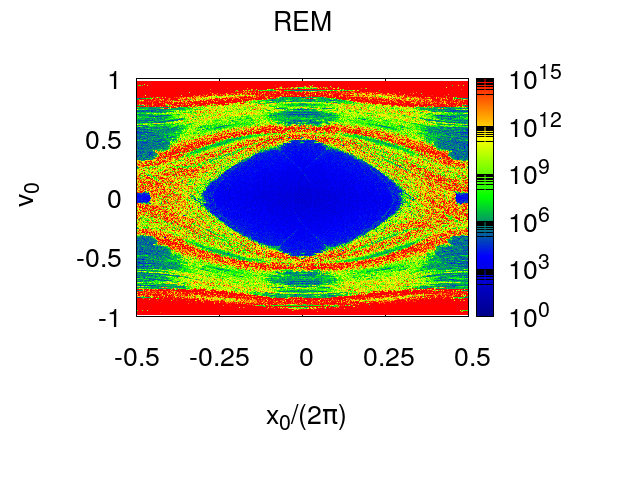}
    \caption{}
\end{subfigure}
\caption{ From the top: (a) Plot of LE for the 4D reflection map  of the 3D waveguide with corrugation amplitude $\eps=0.1$. The initial conditions $(x_0,y_0,v_{x\,0},v_{y\,0})$ where $v_{x\,0}=v_0 \cos\phi_0$ and $v_{y\,0}=v_0 \sin\phi_0$,
 are  chosen in a $2D$ plane obtained by keeping  fixed $y_0=\pi/4$ and $\phi_0=0$. The error LE is computed for $N=200$ iterations
 of the map. We let $(x_0,v_0)$  vary on  a regular grid  of $N_g\times N_g$ points, with $N_g=200$,
 chosen in the the  rectangle  $[-\pi,\pi]\times [-1,1]$. 
 (b) The same plot of RE.  (c) The same plot of REM.   }
\label{fig:fig6} 
\end{figure}

\begin{figure}[!ht]
\centering
\begin{subfigure}{\columnwidth}
    \includegraphics[width=\textwidth]{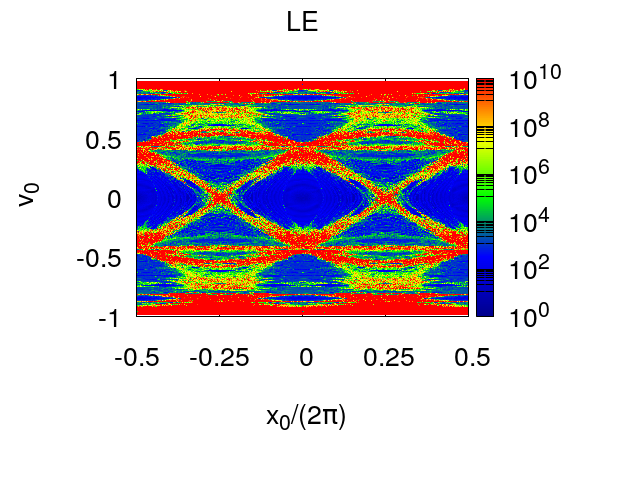}
    \caption{}
\end{subfigure}
\begin{subfigure}{\columnwidth}
    \includegraphics[width=\textwidth]{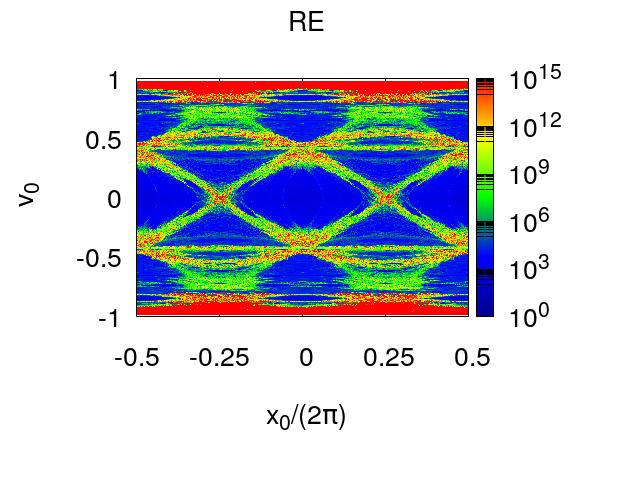}
    \caption{}
\end{subfigure}
\begin{subfigure}{\columnwidth}
    \includegraphics[width=\textwidth]{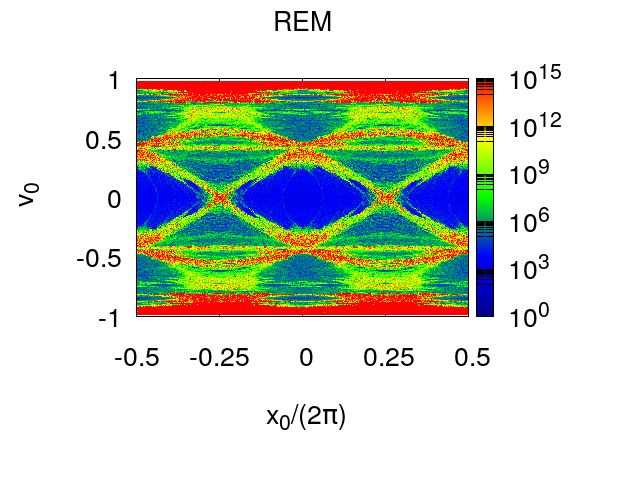}
    \caption{}
\end{subfigure}
\caption{Same plots as figure \ref{fig:fig6} for  LE, RE, REM for $\eps=0.1$ and
  fixed  values of  $y_0=\pi/2$ and $\phi_0=0$.  The number of iterations and grid
  points are the same $N=200,\,N_g=200$. }
\label{fig:fig7} 
\end{figure}

\begin{figure}[!ht]
\centering
\begin{subfigure}{\columnwidth}
    \includegraphics[width=\textwidth]{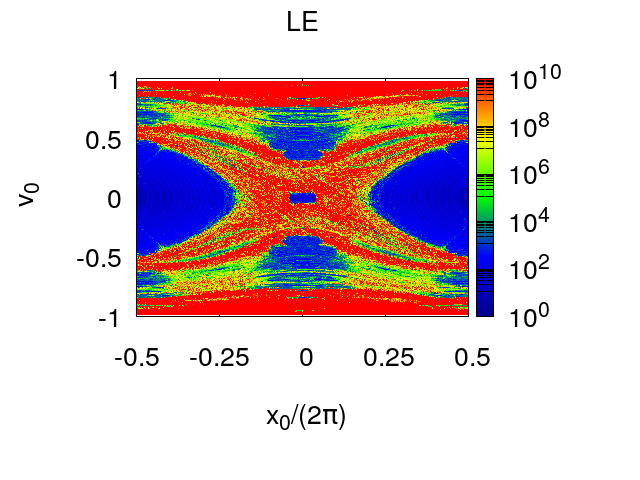}
    \caption{}
\end{subfigure}
\begin{subfigure}{\columnwidth}
    \includegraphics[width=\textwidth]{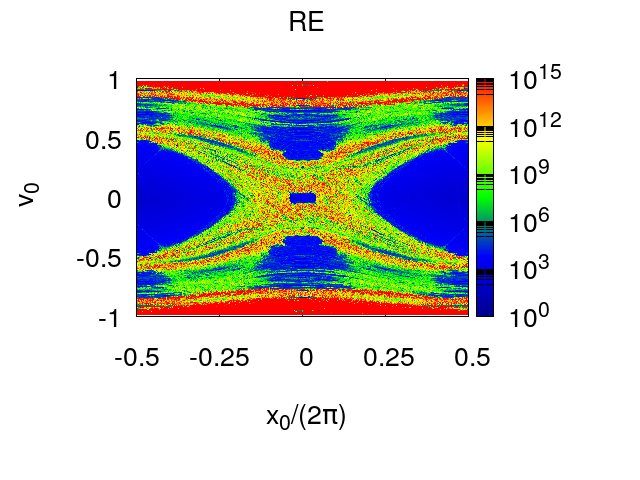}
    \caption{}
\end{subfigure}
\begin{subfigure}{\columnwidth}
    \includegraphics[width=\textwidth]{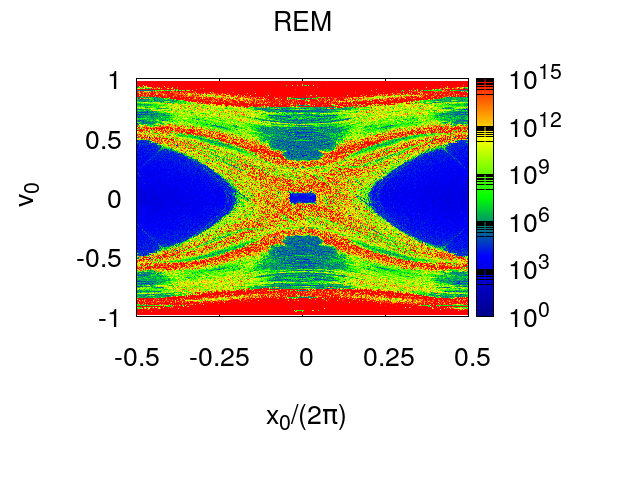}
    \caption{}
\end{subfigure}
\caption{Same plots as figure \ref{fig:fig6} for  LE, RE, REM for $\eps=0.1$ and
  fixed  values of  $y_0=3\,\pi/4$ and $\phi_0=0$.  The number of iterations and grid
  points are the same $N=200,\,N_g=200$. }
\label{fig:fig8} 
\end{figure}

\begin{figure}[!ht]
\centering
\begin{subfigure}{\columnwidth}
    \includegraphics[width=\textwidth]{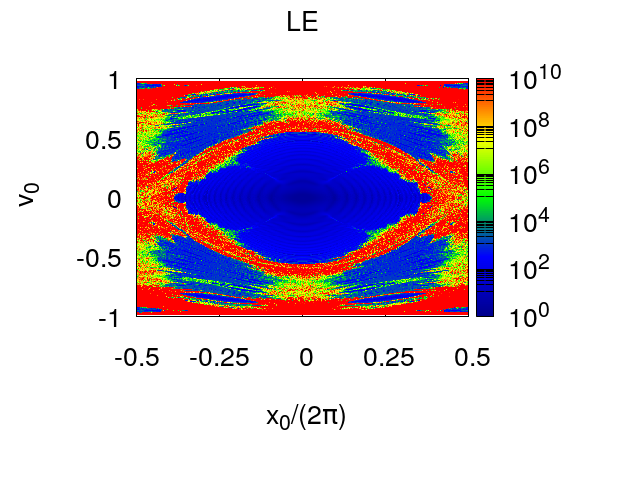}
    \caption{}
\end{subfigure}
\begin{subfigure}{\columnwidth}
    \includegraphics[width=\textwidth]{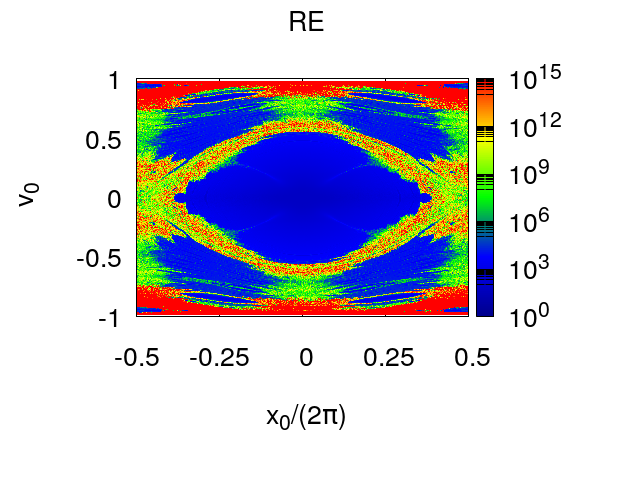}
    \caption{}
\end{subfigure}
\begin{subfigure}{\columnwidth}
    \includegraphics[width=\textwidth]{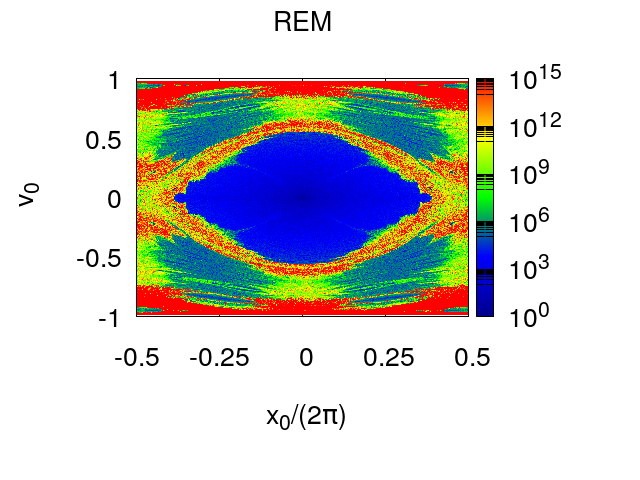}
    \caption{}
\end{subfigure}
\caption{Same plots as figure \ref{fig:fig4} for  LE, RE, REM for $\eps=0.1$ and
  fixed  values of  $y_0=0$ and $\phi_0=\pi/2.01$.  The number of iterations and grid
  points are the same $N=200,\,N_g=200$. }
\label{fig:fig9} 
\end{figure}

For the 3D waveguide  we choose the corrugated profile according to
\begin{equation}
  f(x,y)= \cos x\,\,\cos y ={1\over 2}\cos(x+y)+  {1\over 2}\cos(x-y)
\label{eq31}
\end{equation}
We remark first that  for initial conditions  $y_0=v_{y\,0}=0$ we are back to the 2D case.
If we choose $x_0=y_0$ and $v_{x\,0}=v_{y\,0}$ the ray propagates in the
$x=y$ plane as for 2D waveguide. After a rotation of $\pi/4$
the ray propagates on the $x',z$
plane where $x'=\sqrt{2}\,\,x,\,\,v_x'=\sqrt{2}\,\,v_x$,  the 
corrugation function  depends only on $x'$ as 
$f= {1\over 2}\bigl[\cos(\sqrt{2\,}\,x'\,)+1\bigr]$  and its period
is $\sqrt{2}\,\,\pi$. 
\spa
When the ray does not propagate in a plane the simplest case  to analyze
corresponds to the almost vertical propagation of the ray close to to a
critical point  $(x_c,y_c)$  of the corrugation function, where $\hbox{grad} f=0$.
The fixed point of the map 
$(x_c,y_c,v_x=0,\,v_y=0)$ is  elliptic   if  $f$ has a maximum at
$(x_c,y_c)$   and the corrugated surface near $(x_c,y_c)$ behaves as a concave mirror.
\\
If  $f$ has a minimum at $(x_c,\,y_)$  then the surface behaves as a convex  mirror
and and the critical  point  of the map is hyperbolic.
For our model  near $x=y=0$ we have $f\simeq 1-{1\over 2}(x^2+y^2)$.
The linearized map  becomes 
\begin{equation}
  \begin{split}
    v_{x\,n+1} &=  v_{x\,n}  -2\eps\,(x_n+\tau v_{x\,n} )   \\
    x_{n+1} &= x_n + \tau(v_{x\,n}+v_{x\,n+1}) \\  
    v_{y\,n+1} &=  v_{y\,n}  -2\eps\,(y_n+\tau v_{y\,n} )   \\
    y_{n+1} &= y_n + \tau(v_{y\,n}+v_{y\,n+1})
  \end{split}  
   \label{eq32}
\end{equation}
where $\tau=1+\eps$. Since $\tau$ is constant in this approximation the maps in the $(x,v_x)$ and
$(y,v_y)$ phase planes decouple and each of them is area preserving.
The linear frequencies are equal and given by $\sin^2(\omega/2)=\eps\,\tau$ as for the 2D case.
The degeneracy can be removed choosing for instance $f(x,y)=\cos x+ A \cos y$.
\\
Near $x=0, y=\pi$ we have  $f\simeq 1+{1\over 2}(x^2+(y-\pi)^2)$ and the fixed point
of the map is hyperbolic.  More generally $(0,0),\,(\pm \pi,\,\pm \pi),\,(\pm \pi,\,\mp \pi)$
are elliptic and $(0, \pm \pi),\,(\pm \pi,0)$ are hyperbolic. 
We do not examine the projections of individual orbits on 2D phase space planes
nor on 3D hyperplanes.  Close to the elliptic 
fixed points  single  2D tori are recognisable in  the projections on 3D hyperplanes and 
even 2D phase space planes,
but  the projection of several orbits  does not  provide any useful information.
For this reason we show here the plots of our short term indicators LE, RE and REM
of orbits whose initial points belong to 2D phase planes.
\spa
The  plots are obtained  by computing our indicators for initial
conditions  $(x_0,v_{x\,0},y_0,v_{y\,0})$ in a family of 2D phase planes
where  $y_0$  and $v_{y\,0}/v_{x\,0}$ are kept fixed.  Letting $v_{x\,0}= v_0 \cos(\phi_0)$
and $v_{y\,0}= v_0 \sin(\phi_0)$ we  compare the errors for fixed values of $y_0,\,\phi_0$
letting  $x_0,v_0$ which vary in $[-\pi,\,\pi]\times [-1,1]$, where  we 
choose  a regular  grid of $N_g\times N_g$ points.
We have fixed the corrugation amplitude at
$\eps=0.1$ because there  is a good balance between  regular and chaotic regions.
In figures  \ref{fig:fig6}, \ref{fig:fig7},  \ref{fig:fig8}
we compare the errors LE, RE, REM
in a logarithmic color scale  by choosing  and $y_0=0$  $\phi_0= \,\pi/4, \,\pi/2,\, 3\pi/4$.
We have computed LE by using the shadow orbit to evaluate  the tangent map, namely we have
replaced the partial derivatives $\partial M_i/\partial x_j$ by finite differences, 
avoiding the cumbersome analytic evaluation.  
 To this and we have chosen four different initial conditions $\xbf_0+ \ebf_j\,\,\delta $
for $\le j\le 4$  where $\ebf_j$ are the  orthonormal base vectors in $\Reali^4$ namely
$(\ebf_j)_k=\delta_{j_k}$.
The tangent map is approximated  by
\begin{equation}
  DM^n(\xbf_0)\ebf_j \simeq  \wbf_j(n) \equiv   { M^n(\xbf_0 + \ebf_j\,\delta )- M^n(\xbf_0) \over \delta }
   \label{eq33}
\end{equation}
so that the Lyapunov error becomes 
\begin{equation}
 E^2_{L\,n}\simeq \sum_{j=1}^4 \,\Vert \wbf_j(n)\Vert^2
  \label{eq34}
\end{equation}
the discrepancy  is of order $\delta$ and the choice $\delta=10^{-14}$ was made, using double
precision accuracy.
\spa
For $\phi_0=0$ we recover  the plot computed with the tangent map
for the 2D waveguide since the ray propagates in the $x,z$ plane,   but only if the sum in
equation (\ref{eq34}) runs  only up to 2 (corresponding to the 
phase space coordinates  to $x,v_x$).
This is equivalent to define LE
by replacing $DM^n$ with its first  $2\times2$ block,
which corresponds to tangent map  for the 2D map.
Indeed even though the orbits of the 2D and 4D map are the same for initial conditions
$y_0=v_{y\,0}=0$, the components of 4D tangent map, not belonging the $2\times 2$ block, 
are non zero. On the contrary REM does not change with respect to 2D map, because the
$(x,\,v_x)$ plane is invariant.  To recover agreement between REM and  RE, when LE
is computed according to (\ref{eq34}), which  corresponds to the trace of $(DM^n)^T\,DM^n$, 
one  can add a small random displacement of amplitude $\delta$ before reversing the orbit,
in order to bring the orbit out of the invariant plane. When  there is no invariant
plane the agreement between REM and RE is recovered without any random kick.
%
In figure (\ref{fig:fig9}) (a)-(c) we show the plot for $\phi\simeq \pi/2$
and $y_0=0 $.
The regions of regular motion correspond to neighbourhood of  elliptic points of the resonant
structures,
the regions of chaotic motion to to the separatrices joining the hyperbolic points.
In addition the double resonances, due to single resonances intersection in action space,
create other chaotic regions.  The Fourier analysis might be used to classify the resonances
because the resonant perturbation theory cannot be easily  developed.
\section{Shannon channel capacity}
\begin{figure}[!ht]
\centering
\includegraphics[width=6 cm, height=6 cm]{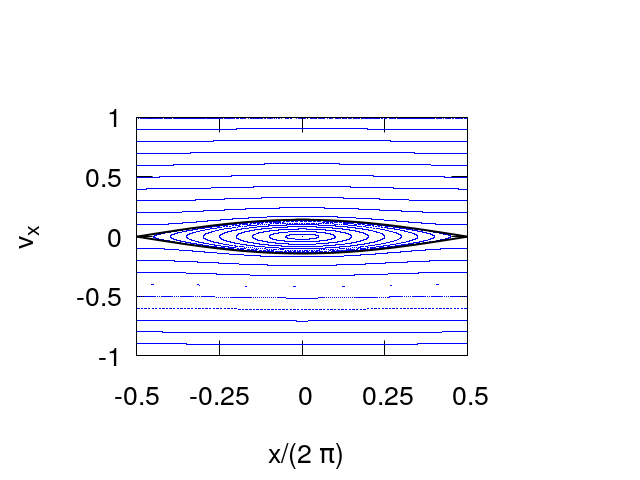}
\includegraphics[width=6 cm, height=6 cm]{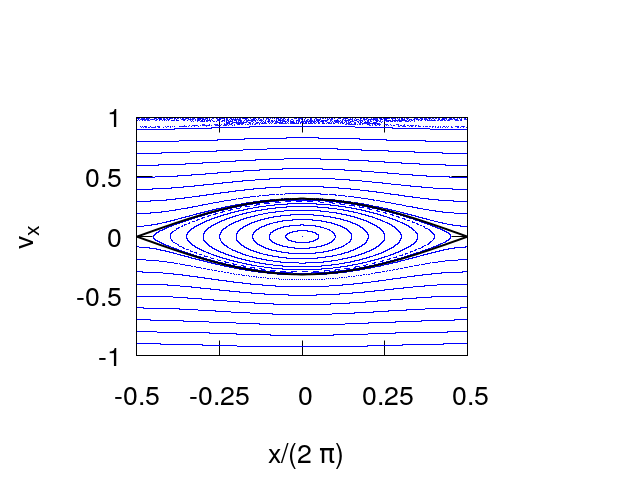}
\includegraphics[width=6 cm, height=6 cm]{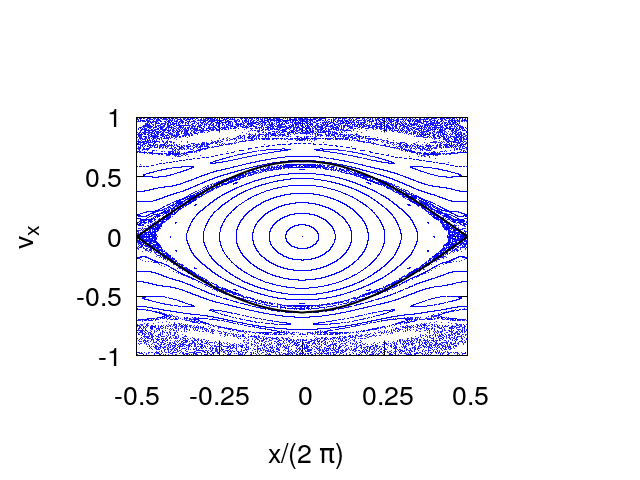}
  \caption{Phase space portraits of the 2D map with the separatrix, black line, given by equation
    (\ref{eq40}), for different values of the corrugation amplitude. Top frame: $\eps=0.005$.
  Center frame: $\eps=0.025$. Bottom frame: $\eps=0.1$.}
\label{fig:fig11} 
\end{figure}

\begin{figure}[!ht]
\centering
\includegraphics[width=6 cm, height=6 cm]{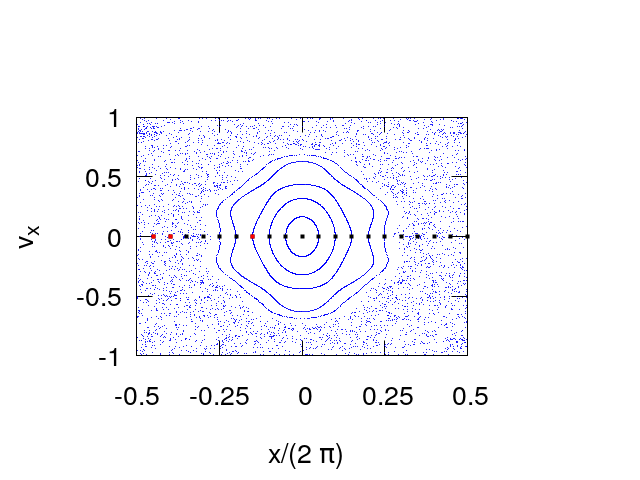}
\includegraphics[width=6 cm, height=6 cm]{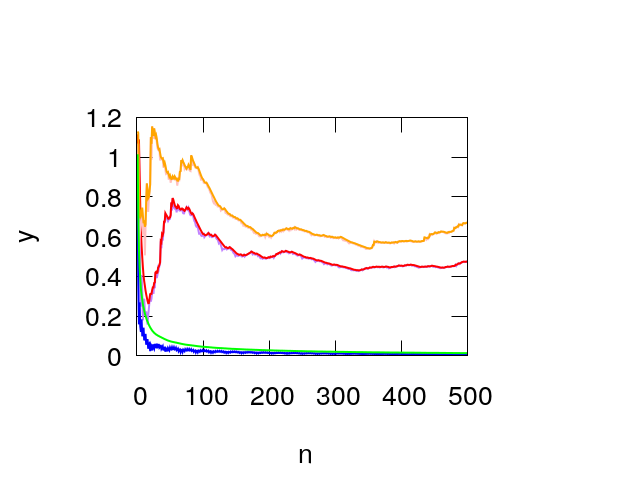}
\caption{Top panel: phase portrait of the 2D ray map for $\eps=0.25$. The blue and red dots correspond
  to the initial conditions. Bottom panel: dependence on $n$ of $C_n^L$  and $C_n^R$ for
  the chaotic orbits  whose initial points are the first two  red dots in left panel (LE pink, RE orange for
  the first orbit, LE purple, RE red for the second orbit)  and the regular orbit whose initial point
is the last red point (LE  blue, RE green) }
\label{fig:fig12} 
\end{figure}

\begin{figure}[!ht]
\centering
\includegraphics[width=10 cm, height=8 cm]{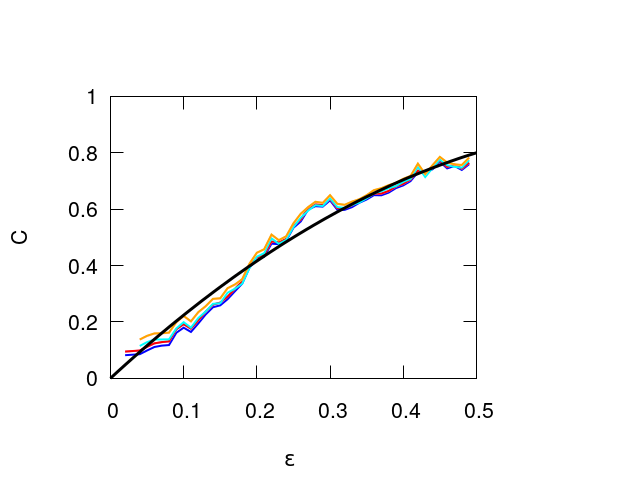}  
\caption{Growth of  the  average channel capacities (on a regular grid of 400 points)
  with $\eps$ for two values of $n$:
 $C_{L\,n }$  cyan for $n=100$, blue for $n=200$ and $C_{R\,n}$, orange for $n=100$, red for $n=200$.
The black line is a  quadratic fit given by equation (\ref{eq41}) }
\label{fig:fig13} 
\end{figure}

The possibility of computing  fast stability indicators allows to establish 
interesting properties that depend on them. 
Among others, in information theory it is important to introduce  the so called Shannon-Hartley channel capacity
\begin{equation}
  C = \lim_{t \rightarrow \infty} \, \frac{1}{t} \, \log_{2} \, \left ( \frac{Y}{X} \right ),
  \label{eq35}
\end{equation}
where the time-dependent relation between input ($X$) and output ($Y$) 
describes the time evolution of the transfer process taking the message from transmitter to receiver. 
Equation (\ref{eq40}) gives the maximum (ideal) information bit data-rate achievable in a physical system that carry 
electric signals/electromagnetic waves. 
In the high-frequency asymptotics, the propagation of waves through the corrugated channel in Fig. \ref{fig:fig1}  
has an isomorphism with classical ray trajectories underlying waves that bounce around the channel, 
given by the symplectic map (\ref{eq9}). 
Therefore, one can think that the maximum Lyapunov exponent has an isomorphism with the Shannon-Hartley channel capacity. 
This has been recognised in \cite{Friedland_2018} to be valid for any dynamical system, hence for a continuous time
system. The relation between the channel capacity and the Lyapunov exponents of random matrices was  first proposed by
\cite{Channel-Capacity}.
\begin{equation}
   C = \lambda_1 = \lim_{t\to \infty}\,{1\over t}\,\log E_L(t) = \lim_{t\to \infty}\,{1\over t}\,\log E_R(t) 
  \label{eq36}
\end{equation}
For recent works on the subject relating channel capacity to Lyapunov exponents and entropy
see \cite{Channel-Entropy}  and \cite{Information-Lyapunov}
Consequently, fast dynamical indicators \cite{Panichi2015,Panichi2017} allow an estimation of the channel capacity. 
For a  map like the waveguide map depending on one parameter $\epsilon$,
given a finite number of iterations $n$ we introduce the sequences 
\begin{equation}
\begin{split}
  C_{L\, n}(\xbf,\eps) &= {1\over n}\,\log E_{L\,n}(\xbf,\eps)  \\
  C_{R\, n}(\xbf,\eps) &= {1\over n}\,\log E_{R\,n}(\xbf,\eps) 
  \end{split}
  \label{eq37}
\end{equation}
having the same limit $C(\xbf,\eps)$ as $n\to \infty$. The  channel capacities
depend on the initial condition $\xbf$,  however if $\xbf$ belongs to an   ergodic component 
the result is the same for almost any choice of $\xbf$.
For a parallel wave guide $\eps=0$  the channel capacities $C_{L\, n}(\xbf,\eps)$ and
$C_{L\, n}(\xbf,\eps)$ vanish asymptotically according to 
\begin{equation}
\begin{split}
  C_{L\,n}(\xbf,\eps) &= {1\over n}\,\log n +O(n^{-1})  \\ 
  C_{R\,n}(\xbf,\eps)  &= {3\over 2n}\,\log n + O(n^{-1})
\end{split}
  \label{eq38}
\end{equation}
For a   corrugated waveguide  the phase plots of $C_{L\,n}(\xbf,\eps)$ and $ C_{R\,n}(\xbf,\eps)$
for fixed $n$ of  are the same as the  plot of  $E_{L\,n}$ and $E_{R\,n}$  in a logarithmic
scale  shown in the  figures \ref{fig:fig3} and \ref{fig:fig4}   for the 2D wave guide and
\ref{fig:fig6} to  \ref{fig:fig9}  for the 4D waveguide, up to the  constant  factor $n^{-1}$.
\spa
The 2D ray reflection map  for $\eps\ll 1$  is almost integrable 
and exhibits two distinct regions  delimited by a separatrix which  can be  approximated by 
\begin{equation}
  v_x= \pm 2\,\sqrt{\eps} \,\cos\parton{x\over 2}
  \label{eq39}
\end{equation}
see  figure \ref{fig:fig11}.
%
%
 For $\eps\sim 0.1$ the separatrix becomes a thin stochastic layer, whose 
area grows  by further  increasing  $\epsilon$.
We have compared the dependence on $n$ of the   channel capacities  $C_{L\,n}(\xbf,\eps)$ and $C_{R\,n}(\xbf,\eps)$ 
for  one  regular  and two chaotic orbits, see  figure  \ref{fig:fig12}.
For the first orbit the channel capacity vanishes as $n^{-1}\,\log n$ whereas for the last two chaotic orbits
a finite limit is approached.
\\
In order to appreciate how the limit for $n\to \infty$
is reached and the dependence on the corrugated amplitude we compute the following phase space averages
\begin{equation}
  \begin{split}
    C_{L\,n}(\eps) &={1\over \mu_L(\Ecal)}\, \int_\Ecal \,   C_{L\,n}(\xbf,\eps)\,d\xbf \\
    C_{R\,n}(\eps) &={1\over \mu_L(\Ecal)}\, \int_\Ecal \,   C_{R\,n}(\xbf,\eps)\,d\xbf
  \end{split}
  \label{eq40}
\end{equation}
where $\Ecal$ denotes the phase space and $\mu_L(\Ecal)$ its volume.
\\
We have analyzed the dependence on the corrugation amplitude $\eps$ of the average channel capacities
$C_{L\,n}(\eps)$  and $C_{R\,n}(\eps)$  for $n=100$ and $n=200$, see  
Figure  \ref{fig:fig13}. 
The channel capacity   varies almost monotonically  from 0 to   0.8   and 
corresponds to the phase space  average  of the maximum Lyapunov exponent, since for $n=200$
the asymptotic value appears to be reached.
We have found the following quadratic fit with $\eps$ 
\begin{equation}
  C(\eps)= 2.4\,\eps -1.6\,\eps^2  \qquad \qquad 0\le \eps\le{1\over 2}
  \label{eq41}
\end{equation}
A similar analysis can be performed for the 4D map  by computing  the phase space average 
with a Monte Carlo sampling rather than on a regular grid as for the 2D map.
%
%
%
%
%
%

%
%
%
\section{Conclusions}
We have analyzed the propagation of a ray on   2D and 3D   waveguide,
given by two parallel lines or planes, 
one of which   has a  periodic corrugation.
In both cases the ray reflection maps  on the uncorrugated line or plane are
symplectic.  The stability properties are relevant for the long term
propagation of the rays. Indeed in the chaotic regions a  diffusion
occurs preventing the coherent propagation of a signal.
We have analyzed the reflection maps using  the short time indicators recently proposed:
the Lyapunov error LE,  and the reversibility errors RE and REM. The  square of LE is 
the trace of  the covariance matrix  of the displacement, after $n$ iterations
of the map, induced by a small initial random
displacement.  The square of RE  is the trace of
the covariance matrix  of  the displacement from  the initial condition,  
when the map is iterated forward and backward $n$ times,  adding  at each step a small
random displacement.
Replacing the random displacement with  the round off allows to define a modified
 reversibility error  REM, which is similar to RE,  though affected by   large fluctuations
due to the absence of averaging.
In the 2D case a good qualitative agreement with the phase
portrait is found. For a fixed number $n$ of reflection of the rays on the uncorrugated plane, 
all the indicators exhibit a similar behaviour with a power law growth  for regular orbits and
exponential growth for chaotic orbits.   For small corrugations the motion is regular
almost everywhere except in the neighborhood of the separatrix, occurring when the rays hit
a minimum of the corrugated line or surface,  and for rays  propagating  almost parallel to the waveguide
\\
For the orbits of the 4D map, describing the ray dynamics of the 3D waveguide,  LE and RE
provide a satisfactory stability portrait, just as REM.   A stability analysis in this case
cannot be performed by looking at the projection of orbits on a 2D phase plane.
Altogether these indicators quantify separately the effect of small initial
displacements, additive noise and round off.  The limited computational load
and the simplicity of implementation of these indicators make  them
a convenient tool  also for a parametric study of 2D and 3D waveguides.
We have also considered the  relation of our indicators with the channel capacity $C$.
Indeed   the limit of $n^{-1}\,\log E_{L\,n}$ and $n^{-1}\,\log E_{R\,n}$
 for $n\to \infty$  gives the maximum Lyapunov exponent which is isomorphic
to the channel capacity. The dependence on $n$ for a given initial condition is consistent
with theoretical estimates.  The phase space average of  the  channel capacity  
is found to rise with the corrugation amplitude $\epsilon$ following a quadratic law.

\section{Data Availability Statement}
The data that support the findings of this study are openly available in \url{https://github.com/gabrielegradoni/WaveguideStability}.

\begin{acknowledgments}
GG wishes to acknowledge the financial support of The Royal Society under the grant INF$|$R2$|$192066.
\end{acknowledgments}

\def\eps{\epsilon}
\section{Appendix A: tangent map for 2D waveguide}
The equation which determines $\tau(x,v)$ is
\begin{equation}
  \begin{split}
  G(x,v,\tau) &= 1+\eps\,f(x+\tau v)-\tau \,\sqrt{1-v^2} =0     \\
  \derp{\tau}{x} &= -{\derp{G}{x}\over \derp{G}{\tau} }= {\eps\,f'\over \sqrt{1-v^2} -\eps v f'}  \\
   \derp{\tau}{v} &=-{\derp{G}{v}\over \derp{G}{\tau} }= {\eps\,\tau\,f' +\tau\,v/ \sqrt{1-v^2}\over \sqrt{1-v^2} -\eps v f'} 
  \end{split}
  \label{eqA1} 
  \end{equation}
Letting   $v_x=cos\theta\equiv v$ and $v_z=\sin \theta \equiv \sqrt{1-v^2}$
the recurrence which determines the symplectic map $M=(x,v)$  is given by
\begin{equation}
  \begin{split}
    & G(x_n,v_n,\tau) = 0 \\  \\
    & v_{n+1} 
    = v_n-2{ \eps^2{f'}^2\,v_n -\eps f'\,\sqrt{1-v_n^2} \over 1+\eps^2{f'}^2 }  \\ \\
   & x_{n+1} 
    =x_n + \tau v_n +\tau\,{ \sqrt{1-v_n^2} \over \sqrt{1-v_{n+1}^2}}\,v_{n+1}
  \end{split}
  \label{eqA2} 
\end{equation}
The tangent map is given by
\begin{equation}
  \begin{split}
    \derp{v_{n+1}}{x_n} &=  {2\eps \, f'' \over (1+\eps^2\,{ f'}^2)^2 }\,
    \parqua{-2\eps\,v_n\,f'+
      \sqrt{1-v_n^2}\,(1-\eps^2{f'}^2)  } \times \\ &\times \parton{1+v_n\,\derp{\tau}{x_n} } \\   
    \derp{v_{n+1}}{v_n} &= 1- {2\,\eps \,f' \over 1+\eps^2\,{f'}^2 }\,\parton{\eps \,f'+{v_n\over \sqrt{1-v_n^2} }} +\\
    & +   {2\eps \, f'' \over (1+\eps^2\,{ f'}^2)^2 }  \parqua{-2\eps\,v_n\,f'+
      \sqrt{1-v_n^2}\,(1-\eps^2{f'}^2)   } \times  \\ & \times \parton{\tau+v_n\,\derp{\tau}{v_n}} \\ 
     \derp{x_{n+1}}{x_n} &= 1 + \parton{ v_n \,+\,  v_{n+1}  { \sqrt{1-v_n^2}\over \sqrt{1-v^2_{n+1}} }  } \,
     \derp{\tau}{x_n} +\\
     & + \tau { \sqrt{1-v_n^2} \over (1-v_{n+1}^2)^{3/2}}\, \derp{v_{n+1}}{x_n}  \\ 
     \derp{x_{n+1}}{v_n} &= \tau  + \parton{ v_n \,+\,  v_{n+1}  { \sqrt{1-v_n^2}\over \sqrt{1-v^2_{n+1}} }  } \,
     \derp{\tau}{v_n} +\\ 
     & + \tau { \sqrt{1-v_n^2} \over (1-v_{n+1}^2)^{3/2} }\, \derp{v_{n+1}}{v_n}  -\\
     & -\tau\,{v_n\over \sqrt{1-v_n^2}}\,{v_{n+1}\over \sqrt{1-v_{n+1}^2}}
     \label{eqA3}
\end{split}
\end{equation}
In all the previous formulae $f=f(x_n+\tau\,v_n)$  with
$f'=f'(x_n+\tau\,v_n)$ and $f''=f''(x_n+\tau\,v_n)$.
\\
In order to check the limit $\eps \to 0$ we recall that the map is given by
\begin{equation}
\begin{split}
 \tau \,\sqrt{1-v_n^2} &=1 \qquad \quad v_{n+1}= v_n \\ 
 x_{n+1} & =x_n + 2 {v_n\over \sqrt{1-v_n^2}}
 \end{split}
\label{eqA4} 
  \end{equation}
In this case the tangent map is given by
\begin{equation}
 \begin{split}
  \derp{x_{n+1}}{x_n} &=1 \\
  \derp{x_{n+1}}{v_n} &= {2\over (1-v_n^2)^{3/2}} \\
  \derp{v_{n+1}}{x_n} &=0 \\ 
  \derp{v_{n+1}}{v_n} &=1 
  \end{split}
\label{eqA4} 
  \end{equation}
In our example we choose  $f(x)$ periodic of period $2\pi$ and scale the coordinates according
to $x=2\pi\,x'$ so that the new tangent map is given by
\begin{equation}
\begin{split}
  \derp{x'_{n+1}}{x'_n}&=\derp{x_{n+1}}{x_n}    \\ \derp{x'_{n+1}}{v_n}&={1\over 2\pi}\, \derp{x_{n+1}}{v_n} \\
  \derp{v_{n+1}}{x'_n}&= 2\pi\, \derp{v_{n+1}}{x_n}
  \end{split}
\label{eqA5} 
  \end{equation}
  
\section{Appendix B. The 3D wave-guide}
The  equation for the corrugates waveguide is given by equation (\ref{eq12})
where $f(x,y)$ is a periodic function. The normal vector $\nubf$ and the tangent
vectors $\taubf_x,\,\taubf_y$ are given by
\begin{equation}
\begin{split}
  \nubf(x,y)&={ (\eps\,f_x,\,\eps\,f_y,\, -1)^T\over \sqrt{1+\eps^2(f^2_x+f^2_y) }}\\
  \taubf_x&= { (1,\,0,\,  \eps f_x)^T\over \sqrt{1+\eps^2\,f^2_x }}\\ 
  \taubf_x&= { (0, 1,\,  \eps f_y)^T\over \sqrt{1+\eps^2\,f^2_y  }}
 \end{split}
 \label{eqB1}
\end{equation}
where $f_x\equiv \partial f/\partial x$ and $f_y\equiv \partial f/\partial y$.
For an analogy with the 2D case we might introduce the curvilinear abscissas $s_x,s_y$  on the lines
on the corrugated plane  having $y$ fixed and $x$ fixed respectively
\begin{equation}
\begin{split}
s_x(x,y) &= \int_0^x \,\sqrt{1+\eps^2 \,f_x^2(x',y)}\,dx' \\ s_y(x,y) &=
\int_0^y \,\sqrt{1+\eps^2 \,f_y^2(x,y')}\,dy'
\end{split}
\label{eqB2}
\end{equation}
so that, keeping y fixed,  $ds_x= \sqrt{1+\eps^2 \,f_x^2(x,y)}\,dx$ and, keeping x fixed,  $ds_y= \sqrt{1+\eps^2 \,f_x^2(x,y)}\,dy$.
The transformation from $(x,y)$ to $(s_x,s_y)$ is invertible. Letting $P_0=(x_0,y_0,0)$
be a point on the plane and $Q=(x,y,z(x,y))$ a point on the corrugated plane and $P_1=(x_1,y_1,0)$ a new point on the
plane we consider a ray whose path is $P_0,Q,P_1$.  We denote with  $\vbf_0$ and $\vbf$  the velocity of the 
ray directed from $P_0$ to $Q$ and from $Q$ to $P_1$. As for the $2D$ case we keep $P_0$ and $P_1$ fixed
letting $Q$ vary and consider the ray path length $H$ which depends on $Q$
\begin{equation}
  \begin{split}
    H(x,y)&= h(x_0,y_0,x,y) +  h(x_1,y_1,x,y)  \\ \\
    h(x_0,y_0,x,y)& =\sqrt{(x-x_0)^2+(y-y_0)^2 +(1+\eps f(x,y))^2}
 \end{split}
\label{eqB3}
\end{equation}
The velocities $\vbf_0$ and $\vbf$, denoting for brevity  $h_0\equiv h(x_0,y_0,x,y)$ and
$h_1 \equiv h(x_1,y_1,x,y)$, are given by 
\begin{equation}
 \begin{split}
  \vbf_0 &={(x-x_0,\,y-y_0,\, 1+\eps f(x,y))\,^T \over h(x_0,y_0,x,y) }  \\
  \vbf &={(x_1-x_0,\,y_1-y_0,\, 1+\eps f(x,y)\,)^T \over h(x_1,y_1,x,y) } 
  \end{split}
\label{eqB4}
\end{equation}
We first compute the derivatives of $h_0$ and $h_1$ with respect to $x$ and $y$
\begin{equation}
\begin{split}
  \derp{}{x}\,h_0& = {x-x_0+ (1+\eps f)\,\eps\, f_x\over h_0} = \vbf_0 \cdot {\taubf}_x \,
  \sqrt{1+\eps^2\,f_x^2}  \\
  \derp{}{y}\,h_0& = {y-y_0+ (1+\eps f)\,\eps\,f_y\over h_0} = \vbf_0 \cdot {\taubf}_y \,
  \sqrt{1+\eps^2\,f_y^2} \\ \\ 
 \derp{}{x}\,h_1& = {x-x_1+ (1+\eps f)\,\eps\,f_x\over h_1} = -\vbf \cdot {\taubf}_x \,
  \sqrt{1+\eps^2\,f_x^2}  \\
  \derp{}{y}\,h_1& = {y-y_1+ (1+\eps f)\,\eps\,f_y\over h_1} = -\vbf \cdot {\taubf}_y \,
  \sqrt{1+\eps^2\,f_y^2} 
\end{split}
\label{eqB5}
\end{equation}
Choosing $h_0$ and $h_1$ a functions of $s_x,s_y$  rather than $x,y$ one has
$\partial{h_0}/\partial s_\alpha= \vbf_0\cdot \taubf_\alpha$ and 
and $\partial{h_1}/\partial s_\alpha= -\vbf\cdot \taubf_\alpha$ for $\alpha=x,y$.
The function $H$ is stationary when $\vbf_0 \cdot \taubf_x=\vbf_0 \cdot \taubf_x$ and 
$\vbf_0 \cdot \taubf_y=\vbf_0 \cdot \taubf_y$  namely when the projection on the tangent
plane of the incoming and outgoing ray velocities are equal. This corresponds to the reflection
condition since    the normal components of $\vbf_0$ and $\vbf$ are opposite.
The derivatives with respect to $x_0,y_0$ and $x_1,y_1$ are given by 
\begin{equation}
\begin{split}
  \derp{h_0}{x_0}\,  &= {x_0-x\over h_0 } = -v_{x\,0}  \qquad 
  \derp{h_0 }{y_0}\, = {y_0-y\over h_0 } = -v_{y\,0}
 \\ \\ 
  \derp{h_1}{x_1}\,  &= {x_1-x\over h_1 } = v_{x\, 1}  \qquad \quad 
  \derp{h_1}{y_1}\,h_1  = {y_1-y\over h_1 } = -v_{y\, 1} 
\end{split}
\label{eqB6}
\end{equation}             
Finally  letting $(x_*,y_*)$ the point where $H$ is stationary and  which depends
on end points $(x_0,y_0)$ and $(x_1,y_1)$ we introduce the function $F$
defined as the value of $H$ evaluated at the stationary point and compute its differential
\begin{equation}
  \begin{split}
    & F(x_0,y_0,x_1,y_1)  = h(x_0,y_0,\,x_*,y_*)+h(x_1,y_1,\,x_*,y_*)   \\ \\
    & dF   =  -v_{x\,0}\,dx_0  -v_{y\,0}\,dy_0 + v_{x\,1}\,dx_1  +v_{y\,1}\,dy_1
\end{split}
\label{eqB7}
\end{equation}
\
As a consequence $F$ is the generating function of a canonical transformation $M$
which maps $(x_0,v_{x\,0},y_0,v_{y\,0})$ into $(x_1,\,v_{x\,1},y_1,v_{y\,1})$.  Denoting with
$\xbf_n=(x_n,v_{x\,n},y_n,v_{y\,n})^T$ the phase space point reached after $n$
iterations of the symplectic map $M$ , the recurrence from $\xbf_n$ to $\xbf_{n+1}$
is obtained first by computing the  transit time $\tau$  from $P_n=(x_n,y_n)$
to the point $Q=(x_*,y_*)$ on the corrugated surface.  Then we notice  that the   horizontal
plane projections  of the  velocities $\vbf_*= \vbf_n-2\nubf(\vbf_n\cdot\nubf)$ 
and  $\vbf_{n+1}$ of the ray  reflected at $Q$ and 
$P_{n+1}=(x_{n+1},y_{n+1})$ are equal.  Finally  we determine the displacement from  $P_n$
to $P_{n+1}$.
 The equations defining the recurrence from $\xbf_n$ to $\xbf_{n+1}$ are given by (\ref{eq14}).


\clearpage
\newpage
\begin{thebibliography}{10}

  \bibitem{ref0} 
    A.~{Bazzani}, P.~{Freguglia},  and G.~{Turchetti}, {Hamiltonian Analytical Optics and Simulations
      of Betatronic Motion by Optical Devices}, 
in {\em Nonlinear Dynamics and Collective Effects in Particle Beam Physics}, World Scientific, pp. 23-46, (2019).

\bibitem{ref1}
  L.A.Bunimovich  {On the Ergodic Properties of Nowhere Dispersing Billiards},
  {\em Commun Math Phys.} {\bf 65 (3)}, 295-312, (1979).

\bibitem{ref2} J. Redmond, S. Tabachnikov {\em }{\em Introducing symplectic billiards}
  \hfill \\
  https://www.math.psu.edu/tabachni/prints/Notes5.pdf

\bibitem{ref3} S. Tabachnikov {\em }{\em  billards} \hfill \\
    https://www.math.psu.edu/tabachni/Books/billiardsbook.pdf
    
\bibitem{ref4} S. Woo Park  {\em An introduction to dynamical billiards } \hfill \\
    https://math.uchicago.edu/~may/REU2014/REUPapers/Park.pdf

\bibitem{ref5}
D.~{Holm} and G.~{Kovacic}, {Homoclinic chaos for ray optics in a fiber },
{\em Physica D} {\bf 51}, p. 177  (1991).

\bibitem{ref6}
S.~S. {Abdullaev} and G.~M. {Zavlaskii}, {Classical nonlinear dynamics and
  chaos of rays in problems of wave propagation in inhomogeneous media}, {\em
  Usp. Fiz. Nauk.} {\bf 161}, p.~1  (1991).

\bibitem{ref7}
D.~{Douglas}, {Chaotic billiard lasers }, {\em Nature} {\bf 465}, p. 696
  (2010).




\bibitem{ref8}
S.~{Creagh}, {Directional Emission from Weakly Eccentric Resonators}, {\em
  Phys. Rev. Lett.} {\bf 98}, p. 153901  (2007).

\bibitem{ref9}
G.~{Tanner}, {Dynamical energy analysis Determining wave energy distributions
  in vibro-acoustical structures in the high-frequency regime}, {\em Journal of
  Sound and Vibration} {\bf 98}, p. 153901  (2007).

\bibitem{ref10}
E.~{Leonel}, D.~{da Costa} and C.~{Dettmann}, {Scaling invariance for the
  escape of particles from a periodically corrugated waveguide}, {\em Physics
  Letters A} {\bf 376}, 421  (2012).

\bibitem{ref11}
J.~{de Oliveira}, C.~{Dettmann}, D.~, {da Costa} and E.~{ Leonel}, {Scaling
  invariance of the diffusion coefficient in a family of two-dimensional
  Hamiltonian mappings},{\em Phys. Rev. E} {\bf 87}, p. 062904,  (2013).

\bibitem{ref12}
G.~{Gradoni}, J.-H. {Yeh}, B.~{Xiao}, T.~{Antonsen}, S.~{Anlage} and O.~E.,
  {Predicting the statistics of wave transport through chaotic cavities by the
  random coupling model: A review and recent progress}, {\em Wave Motion} {\bf
  51}, 606  (2014).

\bibitem{ref13} 
G.~{Forte}, F.~{Cecconi} and A.~{Vulpiani}, {Transport and
  fluctuation-dissipation relations in asymptotic and preasymptotic diffusion
  across channels with variable section}, {\em Phys. Rev E} {\bf 90}, p. 062110, (2014).
http://denali.phys.uniroma1.it/~cecconif/MyPapersPDF/gforte\_PRE90.pdf

\bibitem{ref14} 
F.~{Cecconi}, V.~{Blakaj}, G.~{Gradoni} and A.~{Vulpiani}, {Diffusive transport in highly corrugated channels}, 
{\em Phys. Lett. A}, {\bf 383}, pp. 1084-1091, (2018).



  

\bibitem{Froeschle2000b}    
C. ~{Froeschl{\'e}}  and E. ~{Lega}.
On the Structure of Symplectic Mappings. The Fast Lyapunov
  Indicator: a Very Sensitive Tool.
{\em Celestial Mechanics and Dynamical Astronomy}, {\bf 78}, 167--195, 2000.


\bibitem{Froeschle2000a}   
  C. ~{Froeschl{\'e}}, M. ~{Guzzo}   and E. ~{Lega}.
{Graphical Evolution of the Arnold Web: From Order to Chaos}.
{\em Science}, {\bf 289}, 2108--2110, September 2000.
%

\bibitem{Skokos2001}     
C.~{Skokos}.
{Alignment indices: a new, simple method for determining the ordered
  or chaotic nature of orbits}.
{\em Journal of Physics A Mathematical General}, {\bf 34}, 10029--10043,
  2001.

 
\bibitem{Skokos2010} C. Skokos 
{\em  The Lyapunov Characteristic
  Exponents and Their Computation},J.~{Souchay} and R.~{Dvorak} (eds.), Lecture Notes in Physics, Berlin Springer
  Verlag Vol.~790, March 2010.
  
\bibitem{Barrio2016}    
R.~{Barrio}.
Theory and Applications of the Orthogonal Fast Lyapunov Indicator
  (OFLI and OFLI2) Methods.
{\em Chaos Detection and Predictability}, {\bf915}, 55--92, March 2016.

\bibitem{Cincotta2000}    
P. M. ~{Cincotta}  and C. ~{Sim{\'o}}.
Simple tools to study global dynamics in non-axisymmetric galactic
  potentials - I.
{\em A\&AS}, {\bf 147}, 205--228, December 2000.


\bibitem{Cincotta2003}
 P.~M. {Cincotta}, C.~M. {Giordano} and C.~{Sim{\'o}}, {\em Phase space structure of
   multi-dimensional systems by means of the mean exponential growth factor of
 nearby orbits}, {Physica D Nonlinear Phenomena} {\bf 182}, 151 (August  2003).



\bibitem{Sandor2004}
Z.~{Sandor}, B~{.Erdi}, A.~{Szell},  B.~{Funk}.
{\em The relative Lyapunov indicator: an efficient method of chaos detection}, 
{\em Celestial Mechanics and Dynamical Astronomy} {\bf 90}, p. 127-138, (2004).
https://arxiv.org/pdf/1205.0875.pdf

\bibitem{Bountis2008}
Ch.~{Skokos}, T.~{Bountis}, Ch.~{Antonopoulos}. 
{\em Geometrical properties of local dynamics in Hamiltonian
systems: The Generalized Alignment Index (GALI) method},
{\em Physica D} {\bf 231}, p.  3054, (2007).

\bibitem{Skokos20016}
Ch.~{Skokos}, T.~{Manos}. 
{\em The smaller (SALI) and the generalized (GALI)
alignment indices: Efficient methods of chaos detection}, 
{Chaos Detection and Predictability}   Springer Lecture Notes in Physics, 
Editors: Ch. Skokos, Charalampos. G.A.  Gottwald, J.  Laskar  (Eds.).
https://arxiv.org/pdf/1412.7401.pdf

\bibitem{Voglis1998}
C.~{Voglis}, G.{~Contopoulos}. 
{\em  The relative Lyapunov indicator: an efficient method of chaos detection }
{\em Phys. Rev. E}  Vol {\bf 57}, 372-377  (1998)


\bibitem{Voglis1999}.
C.~{Voglis}, G.{~Contopoulos},  C.~{Efthymiopoulos}. 
{\em Detection of ordered and chaotic motion using the dynamical spectra}
{\em Celestial Mechanics and Dynamical Astronomy}  Vol {\bf 73}, 211-220  (1999) 


\bibitem{Maffione2011}
N. P.{~Maffione}, L. A.~{Darriba} · P. M.~{Cincotta} · C. M.~{Giordano},
{\em comparison of different indicators of chaos based on the
deviation vectors. Application to symplectic mappings}
{Celestial Mechanics and Dynamical Astronomy}  {\bf 111}   (2011) 
https://arxiv.org/abs/1108.2196

\bibitem{Maffione2012}
N. P.{~Maffione}, L. A.~{Darriba} · P. M.~{Cincotta} · C. M.~{Giordano},
{\em Comparative study of variational chaos indicators and ODEs  numerical integrators},
{\em International Journal of Bifurcation and Chaos}    World Scientific Publishing Company  (2012)
{\bf 22}, 1230033 (2012)


\bibitem{KSentropy}
 D. {Ruelle}, 
 {\em An inequality for the entropy of differential maps},
{\em Bol. Soc. Bras. Mat.}, {\bf 9}, 83-87, (1978).



\bibitem{Laskar1992}   
J.  ~{Laskar}, C.  ~{Froeschl{\'e}} and A. ~{Celletti}.
The measure of chaos by the numerical analysis of the fundamental
  frequencies. application to the standard mapping.
  {\em Physica D: Nonlinear Phenomena}, {\bf 56}, 253--269, 05 1992.


\bibitem{Pettini1996} L.~Casetti, C.~Clementi, and M.~Pettini, {\em Riemannian theory of Hamiltonian chaos and Lyapunov exponents},
Phys. Rev. {\bf E 54}, 5969-5984 (1996).

\bibitem{Pettini2008} M.~Cerruti-Sola, G.~Ciraolo, M.~Franzosi, M.~Pettini.
{\em Riemannian geometry of Hamiltonian chaos: hints for a general theory.}
Phys. Rev. E Stat. Nonlin. Soft Matter Phys. {\bf 78}, 046205 (2008)

\bibitem{Gottwald2005}  G.~Gottwald, I.~Melbourne, {\em Testing for chaos in deterministic systems
 with noise}, Physica D: Nonlinear Phenomena {\bf 212},  100-110 (2005) 


\bibitem{Panichi2015}
F.~Panichi, L.~Ciotti and G.~Turchetti, {\em Fidelity and reversibility in the
  restricted three body problem},  Communications in Nonlinear Science and
  Numerical Simulation {\bf 35}, 53   (2015).

\bibitem{Panichi2017}
F.~{Panichi}, K.~{Go{\'z}dziewski} and G.~{Turchetti}, {\em The reversibility error
  method (REM): a new, dynamical fast indicator for planetary dynamics}, MNRAS
{\bf 468}, 469 (June 2017).


\bibitem{Turchetti-INTECH}
  F.~{Panichi} and G.~{Turchetti},
Fast indicators of orbital stability: a survey on
Lyapunov and Reversibility errors {\em INTECH}, (2019)
\pan
https://www.intechopen.com/online-first/fast-indicators-for-orbital-stability-a-survey-on-lyapunov-and-reversibility-errors


\bibitem{Turchetti-PhysicaD}
  G.~{Turchetti},  F.~{Panichi} and K. ~{Gozdziewski}
  A complete set of fast  Lyapunov and Reversibility invariant indicators. To be submitted


\bibitem{Faranda2012}
  D.~{Faranda}, M.~F. {Mestre} and G.~{Turchetti},
  Analysis of Round off Errors with Reversibility Test as a Dynamical Indicator,
  {\em International Journal of Bifurcation and Chaos},
  {\bf 22}, p. 1250215 (September 2012).

\bibitem{Zanlungo}
G. ~{Turchetti}, S. ~{Vaienti} and F. ~{Zanlungo}.
Asymptotic distribution of global errors in the numerical
  computations of dynamical systems.
{\em Physica A Statistical Mechanics and its Applications},
{\bf 389}, 4994--5006, (November 2010).


  \bibitem{Hairer2006}
E. ~{Hairer}, C. ~{Lubich} and G. ~{Wanner}.
{\em  Geometric Numerical Integration: Structure-Preserving
  Algorithms for Ordinary Differential Equations; 2nd ed.}
 Springer, Dordrecht, (2006).


\bibitem{Arcidosso}
  G. ~{Turchetti}, F. ~{Panichi}
Birkhoff normal forms and stability indicators for betatronic motion
{\em Proceedings of the NOCE workshop, Arcidosso sept. 2017}, 
S. Di Mitri, S. Chattopadhyay, M. Cornacchia Editors, World Scientific  (2019) 


\bibitem{Oseledet}
Oseledets, V. I. (1968). {\sl A multiplicative ergodic theorem. characteristic
Ljapunov, exponents of dynamical systems.} Trudy Moskovskogo Matematicheskogo
Obshchestva, 19:179-210.
 


   
\bibitem{Friedland_2018}
G.~{Friedland} and A.~{Metere} (eds.),{\em  Isomorphism between Maximum Lyapunov Exponent 
  and Shannon?s Channel Capacity}, {https://arxiv.org/pdf/1706.08638.pdf}, January 2018.


\bibitem{Channel-Capacity} Tim Holliday, Andrea Goldsmith, Peter W Glynn
{\sl Capacity of Finite State Channels Based on Lyapunov Exponents of Random Matrices}
September 2006IEEE Transactions on Information Theory 52(8):3509 - 3532
DOI: 10.1109/TIT.2006.878230   (2006) 

\bibitem{Channel-Entropy} Tim Holliday, Peter Glynn, Andrea Goldsmith
{\sl On Entropy and Lyapunov Exponents for Finite-State Channels}
http://www.yaroslavvb.com/papers/holliday-entropy.pdf

\bibitem{Information-Lyapunov} J. M. Ebeid {\sl Relating information theoretic limits to the Lyapunov
exponents of a dynamical system}
  https://www.ideals.illinois.edu/bitstream/handle/2142/16882/Ebeid\_Hani-James.pdf?sequence=1\&isAllowed=y
  






  

%
\end{thebibliography}
\end{document}